\documentclass[12pt]{article}
\usepackage{epsfig,amssymb}

\newcommand{\bq}{\begin{equation}}
\newcommand{\eq}{\end{equation}}
\newcommand{\bqa}{\begin{eqnarray}}
\newcommand{\eqa}{\end{eqnarray}}
\newcommand{\ben}{\begin{enumerate}}
\newcommand{\een}{\end{enumerate}}
\newcommand{\bc}{\begin{center}}
\newcommand{\ec}{\end{center}}
\newcommand{\bqb}{\begin{eqnarray*}}
\newcommand{\eqb}{\end{eqnarray*}}

\def\lsim{\lesssim}

\def\wtil#1{\widetilde{#1}}

\def\d{\mathrm d}

\def\eg{{\it e.g. }}

\def\etal{{\it et al. }}

\def\swd{s^2_W}

\def\L{ {\cal L }}

\def\pr#1#2#3{ Phys. Rev. ${\bf{#1}}$:#2 (#3)}
\def\prl#1#2#3{ Phys. Rev. Lett. ${\bf{#1}}$:#2 (#3)}
\def\pl#1#2#3{ Phys. Lett. ${\bf{#1}}$:#2 (#3)}

\def\np#1#2#3{ Nucl. Phys. ${\bf{#1}}$:#2 (#3)}

\def\nim#1#2#3{Nucl. Instr. Meth. ${\bf{#1}}$:#2 (#3)}

\def\JHEP#1#2#3{JHEP ${\bf{#1}}$:#2 (#3)}

\def\sjnp#1#2#3{Sov. J. Nucl. Phys. ${\bf{#1}}$:#2 (#3)}
\def\atdat#1#2#3{Atommic Data and Nuclear Data Tables ${\bf{#1}}$:#2 (#3)}

\begin{document}

\begin{titlepage}
\setlength{\parskip}{0.25cm}
\setlength{\baselineskip}{0.25cm}
\begin{flushright}
DO-TH 04/07\\
hep--ph/0409053\\
November   2004.\\
\end{flushright}

\vspace{1.0cm}
\begin{center}
\LARGE
{\bf Electron Spectra in the Ionization of Atoms by Neutrinos}
\vspace{1.5cm}

\large
G.J.\ Gounaris$^a$, E.A.\ Paschos$^b$, and P.I.\ Porfyriadis$^a$ \\
\vspace{0.5cm}
\normalsize
$^a${\it Department of Theoretical Physics, Aristotle University of
Thessaloniki}\\
{\it GR-54006 Thessaloniki, Greece}\\
\vspace{0.4cm}

\normalsize
$^b${\it Universit\"{a}t Dortmund, Institut f\"{u}r Physik,}\\
{\it D-44221 Dortmund, Germany} \\
\vspace{1.0cm}

\end{center}

\begin{abstract}

For neutrinos of ${\cal O}(10{\rm keV})$ energies, their
oscillation lengths are less than a few hundred meters, thereby
suggesting the fascinating idea of oscillation experiments of
small geometrical size. To help evaluating this idea, a formalism is
developed for calculating the neutrino ionization cross
sections for  H as well as the  noble atoms. This formalism is based on the
use of spin-independent atomic wave functions, and should  very accurately describe
the ionization spectra for H, He, Ne and Ar.
The accuracy is considerably reduced   for the Xe case though,
 where  the spin dependence  in the  wave functions is
 non-negligible. Nevertheless, even for Xe the results remain qualitatively
   correct. In all  cases,  the atomic ionizations
cross section per electron is   found to  be smaller
than the neutrino cross section off free electrons, approaching it
from below as the energy increases to the 100 keV region. At the
10-20 keV range though, the atomic binding effects in the cross
sections and the spectra
are very important and increasing with the atomic number. They are
cancelling out though, when total ionization
cross section ratios, like $\nu_\mu/\nu_e$ or
$\bar \nu_\mu/\bar \nu_e$, are considered. \\

PACS numbers:13.15.+g

\end{abstract}
\end{titlepage}


\section{Introduction}

Neutrino interactions have been observed and analyzed at  energies
from  a few MeV (in reactor and solar neutrino experiments) \cite{Reines},
to several hundred GeV at accelerators.
The lower   keV energy range though,  has not attracted much interest up to now,
due to the smaller cross sections there.

Nevertheless, when intense neutrino  fluxes from radioactive
decays become available, the possibility of very interesting
experiments may arise, such as the    testing of
 the solar neutrino oscillations using  terrestrial  experiments of rather
small geometrical size \cite{Giomataris}. To see this, we remark
that recent results from the KamLAND experiment  gives
\cite{KamLAND}
\bq
\Delta m^2_{12}  =  8.2^{+0.3}_{-0.3} \times
10^{-5} ~ {\rm eV}^2 ~~ ,~~   \tan^2 \theta_{12}  =
0.39^{+0.09}_{-0.07} ~~, \label{solar-oscillation}
\eq
which would
suggest a corresponding  neutrino oscillation length of about 15
meters for 1 keV neutrinos, increasing to 150 meters when the
energy reaches the 10 keV level. The  oscillation length
associated to $\Delta m^2_{13}\simeq \Delta m^2_{23}$ is about 27 times
smaller \cite{neutrino-data}.

To perform oscillation measurements at keV energies,
the ionization of atoms by neutrinos may be used,
 taking of course into account the binding of the electrons.
As an example we mention that the realization of an intense Tritium
source  may  allow the
study of  $\bar \nu_e$ oscillations in a room size  experiment.

Such experiments may    also help in investigating    the weak
interaction couplings at very low energies \cite{Giomataris},   the measuring
of  $\sin^2\theta_{13}$   \cite{Vergados}, the
search for  measurable   contributions from a neutrino magnetic
moment   \cite{Giomataris, Dobretsov, Volpe}, or contributions from
possible new gauge bosons \cite{Miranda}. These are interesting
questions, which were never investigated at very low energies, and
may provide useful additional information on neutrino physics.

On the theoretical side, there are three calculations
known to us investigating such  effects.
In \cite{Dobretsov},  the energy spectrum of electrons knocked out
from $^{19}F(Z=9)$ and $^{96}Mo(Z=42)$ atoms is computed,
using  relativistic electron
wave functions. It is found that the electron
ionization spectra differ significantly and  are always smaller   than those of the
 free scattering case.

In   \cite{GPP}, we have also computed the total cross section on free electrons,
as well as on  bound electrons in some light  atoms.
At that time we studied the three atoms H, He and Ne. The result again was  that
the ionization cross section per electron  is always smaller than
the cross section from free electrons \cite{GPP},   and
never larger \cite{Gaponov}. Moreover, this
 ionization cross section always decreases as we proceed  from H to He to Ne,
obviously due to the increase of the binding \cite{GPP}.

For planning and carrying out  these  new experiments though,
it is needed   to have the energy--distribution  of
the ionization electrons  for the  noble  gases He, Ne and possibly Ar.
The reason is that  the detector sensitivity  may
generally depend on the ionization electron energy and  it anyway
diminishes below a certain limit; say \eg $\sim 10 \rm eV$
\cite{Giomataris}.
In principle,  Xe could also be useful,
but such an experiment would be much more expensive than a Ne or Ar
one \cite{Vergados}.
Finally, the use of Kr is   hindered, by the existence of a $\beta^-$-emitting
isotope   \cite{Vergados}.

 The aim of the present paper is to present a method for calculating
  such   spectra. We  restrict
 to cases where it is adequate to consider
 the very accurate spin independent atomic wave
 functions of \cite{wave-functions}.  This way, the most interesting
 cases of the He, Ne and Ar
 atoms are covered, for which the  results should  be quite precise.

In principle our method could  be extended to Xe, for which spin independent
wave functions are  also given in \cite{wave-functions}. But the results
would  then be   less accurate\footnote{For an estimate of their error,
see the discussion at the
 end of Section 2.}. This is  suggested by the observation of
a considerable spin dependence in the measured Xe binding energies
\cite{e-binding}, indicating that a dedicated   relativistic treatment is
needed for it.  Thus only a very brief discussion
of the integrated Xe ionization cross section is offered here.

 Finally, since the properties of the detectors needed to experimentally study
 these ionization spectra are intensively  investigated at present
 \cite{Giomataris, Vergados, Giomataris1},
 the feasibility of this task is conveniently left outside
 the scope of the present work.\\

\begin{figure}[t]
\vspace*{-2cm}
\[
\hspace{-1.cm}\epsfig{file=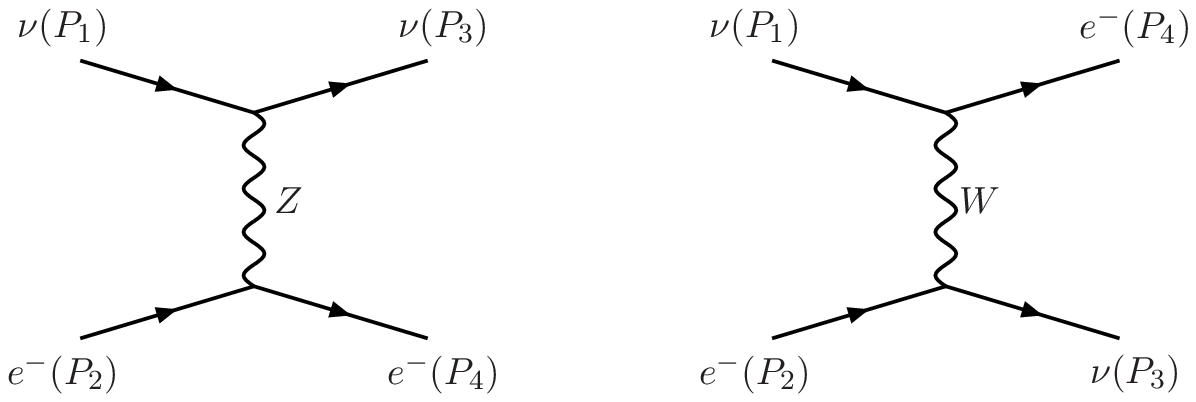,height=3.5cm, width=11.cm}
\]
\vspace*{-1cm}
\caption[1]{The  Feynamn diagrams for $\nu_ee^-$-scattering.}
\label{Feyn-fig}
\end{figure}

At the  very low energies we are interested in, the Standard Model  dynamics
described by the diagrams in Fig.\ref{Feyn-fig} induce
the   effective interaction Lagrangian
\bqa
\L_{e \nu}&= &-\frac{G_F}{\sqrt{2}} \Bigg \{
\Big [ \bar \nu_e \gamma^\rho \frac{(1-\gamma_5)}{2}\nu_e \Big ]
\Big [ v_e \bar e \gamma_\rho e -a_e \bar e \gamma_\rho\gamma_5 e
\Big ] \nonumber \\
&& +\Big [ \bar \nu_\mu \gamma^\rho \frac{(1-\gamma_5)}{2}\nu_\mu
+\bar \nu_\tau \gamma^\rho \frac{(1-\gamma_5)}{2}\nu_\tau \Big ]
\Big [\tilde v_e \bar e \gamma_\rho e -\tilde a_e \bar e \gamma_\rho\gamma_5 e
\Big ] \Bigg \} ~~ , \label{Lagrangian}
\eqa
describing the electron interactions  with any neutrino flavor.
The $\nu_e$ and $(\nu_\mu,\nu_\tau)$ couplings are respectively given by
\bqa
v_e=1+4 \swd &, & a_e=1 ~~,~~ \nonumber \\
\tilde v_e=-1+4 \swd &,&  \tilde a_e=-1~~,
\label{neutrino-couplings}
\eqa
while $G_F$ is  the usual Fermi coupling.

We first concentrate on the  $\nu_e (\bar \nu_e) $ cases.
From (\ref{Lagrangian}),
the squared  invariant amplitude
$|F|^2$, summed over all  initial and final electron spin-states for
the process
\bq
\nu_e ( P_1)~ e^-(P_2)
\to \nu_e (P_3)~ e^-(P_4) ~~  \label{process}
\eq
is calculated,  with the four-momenta being
indicated in parentheses. The various  particle energies
are denoted below by $E_j$, and the standard variables
\bq
 s=(P_1+P_2)^2>m_e^2~~,~~ t=(P_1-P_3)^2~~,~~ u=(P_1-P_4)^2~~,
 \label{Mandelstam}
\eq
 are   used.  We note that (\ref{process}) describes the $\nu_e$
 scattering from either a free or bound electron, the difference
 being determined by $E_2$; and, of course, the
folded in   momentum-wave function in the
 bound electron case. \\

Neglecting neutrino masses, we start from the   case  where
the initial electron is  free,
so that $P_1^2=P_3^2=0$, $P_2^2=P_4^2=m_e^2$, with  $m_e$ being the
electron mass.
Summing over all initial and final electron
spin states, we  then have
\bqa
|F(\nu_e e^- \to \nu_e e^-)|^2_{\rm free} &= &
2 G_F^2 \Big \{ (v_e+a_e)^2 (s-m_e^2)^2
\nonumber \\
&+ & (v_e-a_e)^2 (u-m_e^2)^2
+2 m_e^2 (v_e^2-a_e^2)  t\Big \}
 ~. \label{Fsquared-free}
\eqa
Due to crossing,
 the corresponding $|F|^2$  expression for the $\bar \nu_e$ case
  is simply obtained from the   $\nu_e$ result,
by interchanging $s\leftrightarrow u $ in (\ref{Fsquared-free}).

In the lab system where the initial electron is at rest ($E_2=m_e$),
the differential cross section describing
the energy distribution of the final electron is\footnote{Notice that the
laboratory angle of the emitted electron is not an independent variable, but is fixed
completely by its energy.}
\bqa
\frac{\d \sigma(\nu_e e^- \to \nu_e e^- )}{\d E_4} \Big |_{\rm free}
& =& \frac{m_e G_F^2}{8 \pi E_1^2} \Bigl \{
(v_e+a_e)^2 E_1^2 +   (v_e-a_e)^2 (E_1 +m_e - E_4)^2
 \nonumber \\
&+ & m_e (v_e^2-a_e^2) (m_e-E_4) \Bigr \} ~~ ,
\label{dsigma-nu-free}
\eqa
which, after it is integrated over the allowed range
\bq
m_e \leq  E_4 \leq m_e ~+~ \frac{2 E_1^2}{m_e+2 E_1} ~~ , \label{E4-range-free}
\eq
leads to the $\nu_e $ total cross section off free electrons
\bqa
\sigma_{\rm free}^{\nu e} & \equiv &
\sigma(\nu_e~e^- \to  \nu_e~e^- )\Bigr |_{\rm free}
= \frac{m_e G_F^2 E_1}{8 \pi } \Bigl
\{(v_e+a_e)^2 \frac{2E_1}{m_e+2E_1} \nonumber \\
&+ & \frac{1}{3}(v_e-a_e)^2
\Bigl[1-\frac{m_e^3}{(m_e+2E_1)^3}\Bigr]
-(v_e^2-a_e^2)\frac{2m_e E_1}{(m_e+2E_1)^2} \Bigr \}
~~  . \label{sigma-nu-free}
\eqa

  Starting from  (\ref{dsigma-nu-free}), we present
in\footnote{We always use $\swd=0.23$.}
Fig.\ref{Free-energy-fig} the energy distribution of
the final   electrons,  in $\nu_e$ and $\bar \nu_e$
scattering off free initial $e^-$. In these figures $E_e\equiv E_4$
is the final electron energy,
and $E_\nu$ ($E_{\bar \nu}$) denote the incoming neutrino (antineutrino)
energy $E_1$. As seen there, the energy distribution
always has  maximum at small $E_e$, which is  also seen in
all $\nu_e (\bar \nu_e)$-induced
ionization cases; see below.

In  Section 2 the formalism describing the ionization of atoms
through neutrino (antineutrino) scattering is presented.
This is done first
for the Hydrogen atom,  and it is subsequently generalized to  any  atom
characterized by complete electronic shells. A discussion of the accuracy
of our method for the He, Ne, Ar and Xe atoms is also given.

In Section 3,
we  present the results for the energy distributions of the
knocked out ionization electron in $\nu_e (\bar \nu_e)$ scattering
for  H, as well as for He, and Ne, using  the analytic
spin-independent  Roothann-Hartree-Fock
wave functions published in \cite{wave-functions}.
In addition,   we also present results for  Ne
ionization  through   $\nu_\mu$ or $\nu_\tau$ scattering,
which might be generated through $\nu_e $ oscillation.
Results for the integrated cross sections are also given
in the same Section 3, where we also include our predictions for
the Xe case. The Conclusions are presented
in Section 4.  Important details on the
kinematics and atomic wave functions
are relegated to the Appendices A.1 and A.2 respectively.\\

\section{The formalism.}

The  ionization of  atoms through neutrino
scattering off  atomic electrons, requires from  atomic physics
the electron binding energy and the wave functions in momentum space. The
energy of the initial bound electron is fixed by
the binding energy $\epsilon$, so that
\bq
E_2=m_e+\epsilon ~~,  \label{E2-bound}
\eq
with $\epsilon $ being negative. For the simplest
H atom, $\epsilon$ is fixed by
the Balmer formula, while for more complicated atoms, we use
the experimental measurements \cite{e-binding}.

The  three-dimensional  momentum of the bound electron however,
(compare definitions in (\ref{kinematics})), varies according to
the probability distribution
 \bq
 |\wtil \Psi_{nlm}(p_2, \theta_2, \phi_2) |^2
\frac{p_2^2 \d p_2 \d \cos\theta_2 \d \phi_2 }{(2\pi)^3}~~, \label{probability}
 \eq
 determined by the  momentum  wave
 function\footnote{The momentum here is connected to the
 wave function determining its probability distribution. It is the conjugate
 variable to the position
coordinates, and their wave functions are related through
a Fourier transformation. See Appendix A.2.}
 $\wtil \Psi_{nlm}(p_2, \theta_2, \phi_2)$, when electron spin effects are
neglected. Therefore,
 the bound electron may get off-shell with an effective
 squared-mass given by
 \bq
 \tilde m^2 \equiv P_2^2= E_2^2- p_2^2 ~~, \label{mtilde2}
 \eq
so that, the standard kinematical variables defined in
(\ref{Mandelstam})  satisfy
\bq
s+t+u=\tilde m^2+m_e^2 ~~. \label{Mandelstam1}
\eq

The squared invariant amplitude for the $\nu_e$-electron
subprocess, summed  over all initial and final electron
spin states\footnote{Justified because we neglect any spin dependence on
the wave function.}, is written  as \cite{GPP}
\bqa
|F(\nu_e e^- \to \nu_e e^-)|^2 &= &
2 G_F^2 \Big \{ (v_e+a_e)^2 (s-m_e^2)(s-\tilde m^2)
\nonumber \\
&+ & (v_e-a_e)^2 (u-m_e^2)(u-\tilde m^2)
+2 m_e^2 (v_e^2-a_e^2)  t\Big \}
 ~, \label{Fsquared}
\eqa
which of course becomes identical to (\ref{Fsquared-free}) when
$\tilde m^2 \to m_e^2$. As in the free electron case,
the antineutrino result  may be obtained
from (\ref{Fsquared}) by interchanging $s \leftrightarrow u$.\\

For clarity, we  first consider the ionization of a
Hydrogen atom being   initially   in its ground state.
The energy spectrum of the ionized electron
is obtained  by averaging over the bound electron momenta according
to its wave function (see  Eqs.(15) of \cite{GPP}),
and subsequently changing the $\d \sigma/\d u$-distribution to
$\d \sigma/\d E_4$, which  brings in the derivative $\d u/\d E_4$.
We thus get
\bq
 \frac{ \d \sigma_{\rm H}^{\nu e}}{\d E_4}=
\frac{1}{ 64 \pi   E_1 E_2 } \int
\frac{\d\phi_2 \d\cos\theta_2~ p_2^2 \d p_2}{(2\pi)^3(s-\tilde m^2)}
|\tilde \Psi_{100}(p_2)|^2 |F|^2 \Big |\frac{\d u}{\d E_4}\Big |
~~ . \label{dsigma-dE4-H}
\eq
The variables  $(p_2, \theta_2, \phi_2)$ are the momentum and angles
of the bound electron in the rest frame of the atom,
(compare (\ref{kinematics})); and\footnote{This  wave function has no
 angular dependence, because of the vanishing of
 the orbital angular momentum.}
  $\tilde \Psi_{100}(p_2)$
is the ground state momentum  wave function
defined in (\ref{H-function}). Finally $E_2$ is determined
through (\ref{E2-bound}), which for the H ground  state $(n=1, l=0)$ is given by
\bq
\epsilon =\epsilon^H_{1s} = - \frac{m_e \alpha^2}{2}= -13.6 eV ~~.
\label{H-1s-binding}
\eq

The kinematics are fully explained in Appendix A.1; (compare
(\ref{process})).
According to it, for any $(\theta_2,\phi_2)$ in the range
(\ref{theta2-phi2-range}), and any value of the bound
electron's momentum $p_2$ and the incoming neutrino energy
$E_1$, the polar angle of the ionization electron $\theta_4$ is
a function of its energy $E_4$ given by\footnote{See the discussion
immediately after (\ref{theta4-solution}) for resolving any ambiguity.}
(\ref{theta4-solution}).
This function is  used to determine through  (\ref{du-dE4}),
the expression for $\d u/\d E_4$ needed in (\ref{dsigma-dE4-H})

We also note that the angular dependence of the integrant  in
 (\ref{dsigma-dE4-H}) is  only due to (\ref{du-dE4}) and the
 effect of (\ref{Mandelstam2}) on $|F|^2$.
 The  angular integration is done numerically, with its range
 fixed by (\ref{theta2-phi2-range}). For
the numerical evaluation of the $p_2$-integral, the relevant part of the
$p_2$-range is determined by the form of the electron wave function, as discussed
immediately after (\ref{H-function}) and at the end of Appendix A.2.\\

We next turn to the general case of any of the noble gas atoms He, Ne, Ar,
Kr and Xe. Their  wave functions  are discussed in
Appendix A.2 \cite{wave-functions}.
Since  for noble atoms all electronic shells are complete,
a   summation of the form
\[
\sum_{m=-l}^l |Y_{lm}|^2= \frac{2l+1}{4\pi}~~,
\]
always appears in the wave function contribution to ionization,
washing out any  angular dependence from it.
As a result, the energy distribution
of the ionization electron for any noble atom, normalized to one electron
per unit volume,   may be obtained from (\ref{dsigma-dE4-H}) by
replacing
\bqa
|\tilde \Psi_{100}(p_2)|^2 & \to & \frac{1}{Z\cdot 4\pi} \Bigg \{
2 [\tilde R_{10}(p_2)]^2 + 2 [\tilde R_{20}(p_2)]^2
+ 6 [\tilde R_{21}(p_2)]^2 \nonumber \\
&& ~~~~~~~~ + 2 [\tilde R_{30}(p_2)]^2 + 6 [\tilde R_{31}(p_2)]^2
\nonumber \\
&&~~~~~~~~ + 10 [\tilde R_{32}(p_2)]^2 + 2 [\tilde R_{40}(p_2)]^2
+ 6 [\tilde R_{41}(p_2)]^2 \nonumber \\
&&~~~~~~~~ + 10 [\tilde R_{42}(p_2)]^2 + 2 [\tilde R_{50}(p_2)]^2
+ 6 [\tilde R_{51}(p_2)]^2 \Bigg \}~, \label{He-Xe-atoms}
\eqa
where $\tilde R_{nl}(p_2)$ are the radial momentum wave functions
defined in (\ref{Rtil}), and $Z$ is the atomic number.

Focusing to the expression within the curly brackets in the r.h.s of
(\ref{He-Xe-atoms}),
we  remark that by restricting to just  the first
(one, three, five, eight, all) terms,
 we obtain respectively the results for  the (He, Ne, Ar, Kr, Xe) atoms,
provided the appropriate $Z$ value is used. The corresponding
atomic wave functions are discussed in Appendix A.2 and
\cite{wave-functions}.

We next turn to the discussion of the reliability of our
calculation. We first recall that a very high accuracy of more
than 8 digits has been claimed by the authors of
\cite{wave-functions}, for all RHF (spin averaged) energies of
their results, for all atoms from\footnote{See in particular the
last sentences in Appendix A.2 quoted from their paper.}
 He to Xe.  The main weakness  comes therefore from
the  neglect of spin effects in the atomic wave functions. This
should be   no problem for the lighter atoms up to Kr, for which
the experimental data on the electron binding energies indicate
that the spin effects are  smaller than   $\sim 1\%$ \cite{e-binding}.
 Therefore, our results for He and Ne, and those that could be obtained by
 applying our procedure to   Ar and Kr, should be accurate at the $2\%$-level,
 their main error coming from electromagnetic
 radiative corrections and small relativistic
 spin effects.

For Xe though, the   spin  dependence of the electron binding
energies  shown in Table 1 is  more significant, suggesting that the error in
 the individual  ionization cross section could possibly reach the 10\%
 level\footnote{Note that the stronger singularity of the relativistic
 wave functions at small distances, which increases the  momentum wave
 functions at large momenta, should have no effect on the ionizations,
 since in this range the wave functions are strongly suppressed.}.
In the discussion of Section 3 we find indications though,
that   this error is in fact
diminished  in the ratio of the  $\nu_\mu/\nu_e$ total cross sections
determining $\theta_{13}$. The electron binding effect seems
cancelling in this ratio.
\begin{table}[hbt]
\begin{center}
{ Table 1: Relative spin effect in the  binding energies\\
 of the various electronic quantum states in Xe, \cite{e-binding}. }\\
  \vspace*{0.3cm}
\begin{small}
\begin{tabular}{||c|c||}
\hline \hline
 Quantum  states & Spin effect      \\ \hline
$2p_{1/2}-2p_{3/2}$  & 6.6\%  \\
$3p_{1/2}-3p_{3/2}$ & 6.4\% \\
$3d_{3/2}-3d_{5/2}$ & 1.8\% \\
$4p_{1/2}-4p_{3/2}$ & 0.8\% \\
$4d_{3/2}-4d_{5/2}$ & 2.9\% \\
$5p_{1/2}-5p_{3/2}$  & 10.4\%  \\
\hline \hline
\end{tabular}
 \end{small}
\end{center}
\end{table}

\section{Results for neutrino ionization of Atoms.}

Using (\ref{dsigma-dE4-H}, \ref{H-1s-binding}) and the H-wave function
in (\ref{H-function}), we obtain the results in Fig.\ref{H-energy-fig}
describing the energy distributions of the knocked out electron in
$\nu_e (\bar \nu_e)$ ionization of Hydrogen, normalized to one
electron per unit volume. The results apply  to various incoming neutrino
energies.

\begin{figure}[p]
\vspace*{-1cm}
\vspace*{-1cm}
\[
\hspace{-0.5cm}\epsfig{file=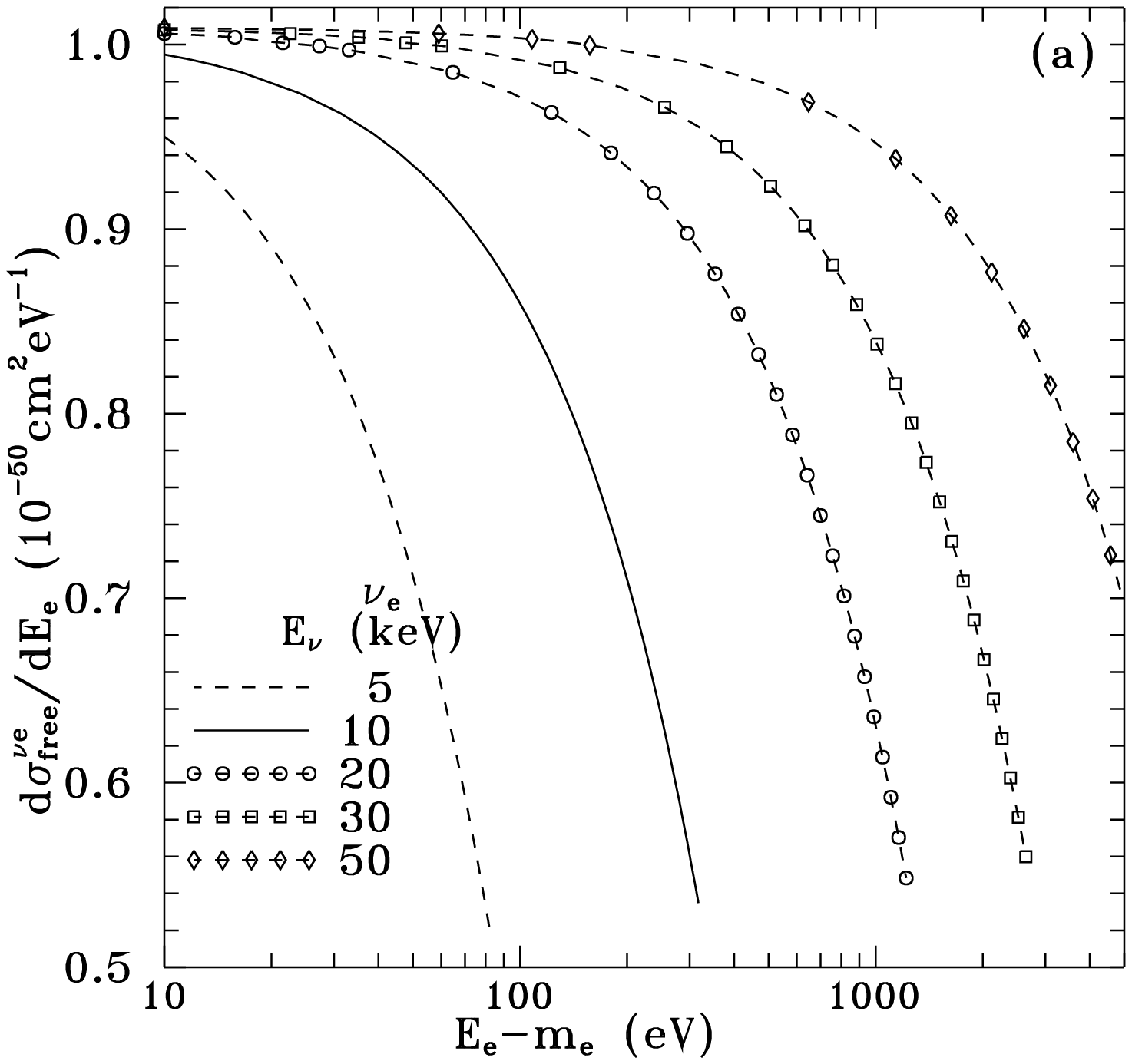,height=7.5cm, width=7.5cm}
\hspace{1.cm}\epsfig{file=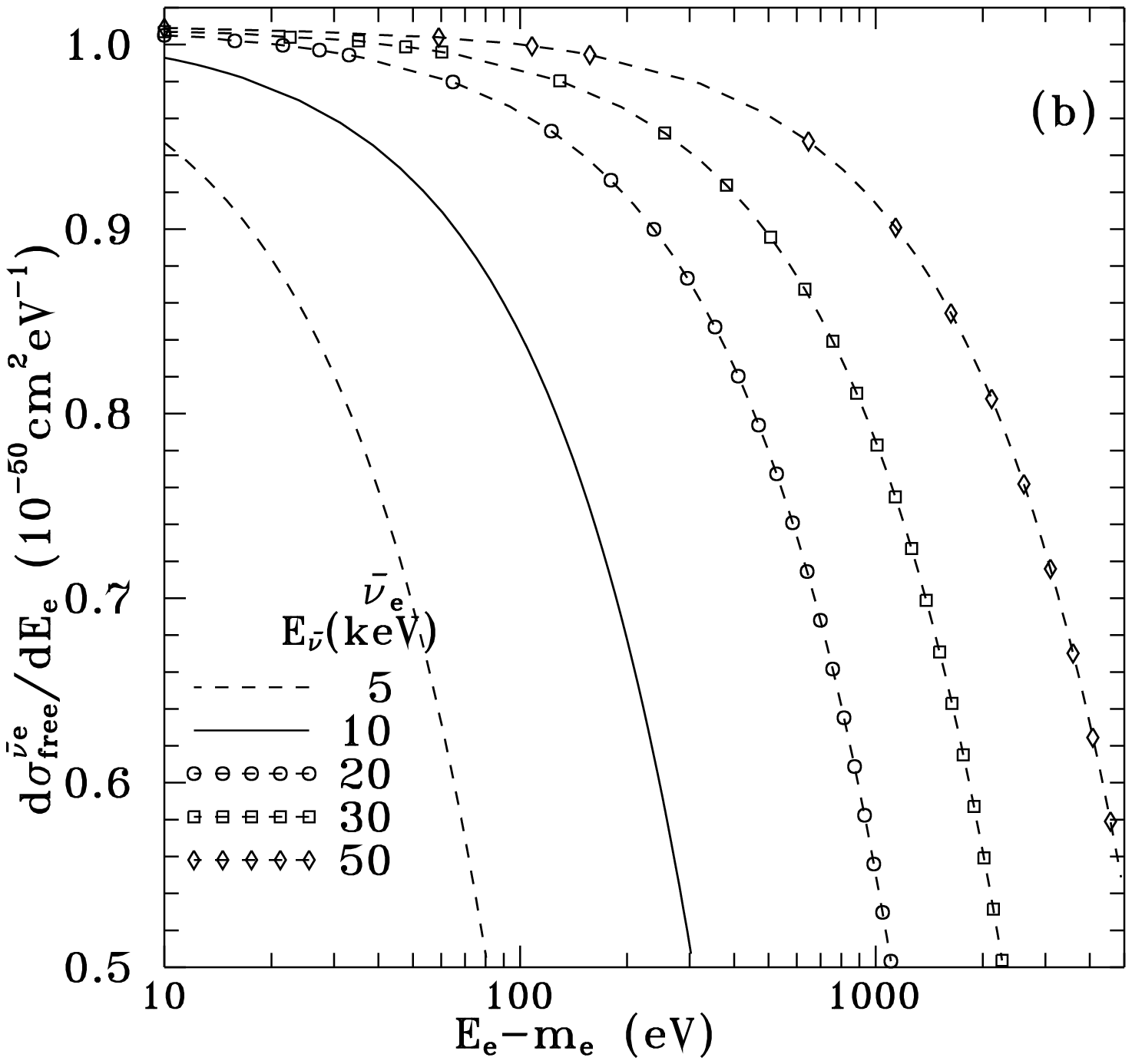,height=7.5cm, width=7.5cm}
\]
\caption[1]{The  energy distribution of the final electron in
$\nu_e e^- $ (a) or $\bar \nu_e e^- $ (b) scattering, at the rest
frame of a free initial $e^-$. Here $E_e\equiv E_4$ is the final
electron energy, and $E_\nu$ ($E_{\bar \nu}$) denote the incoming
neutrino (antineutrino) energy $E_1$; see text.}
\label{Free-energy-fig}
\end{figure}

\begin{figure}[p]
\vspace*{-2cm}
\[
\hspace{-0.5cm}\epsfig{file=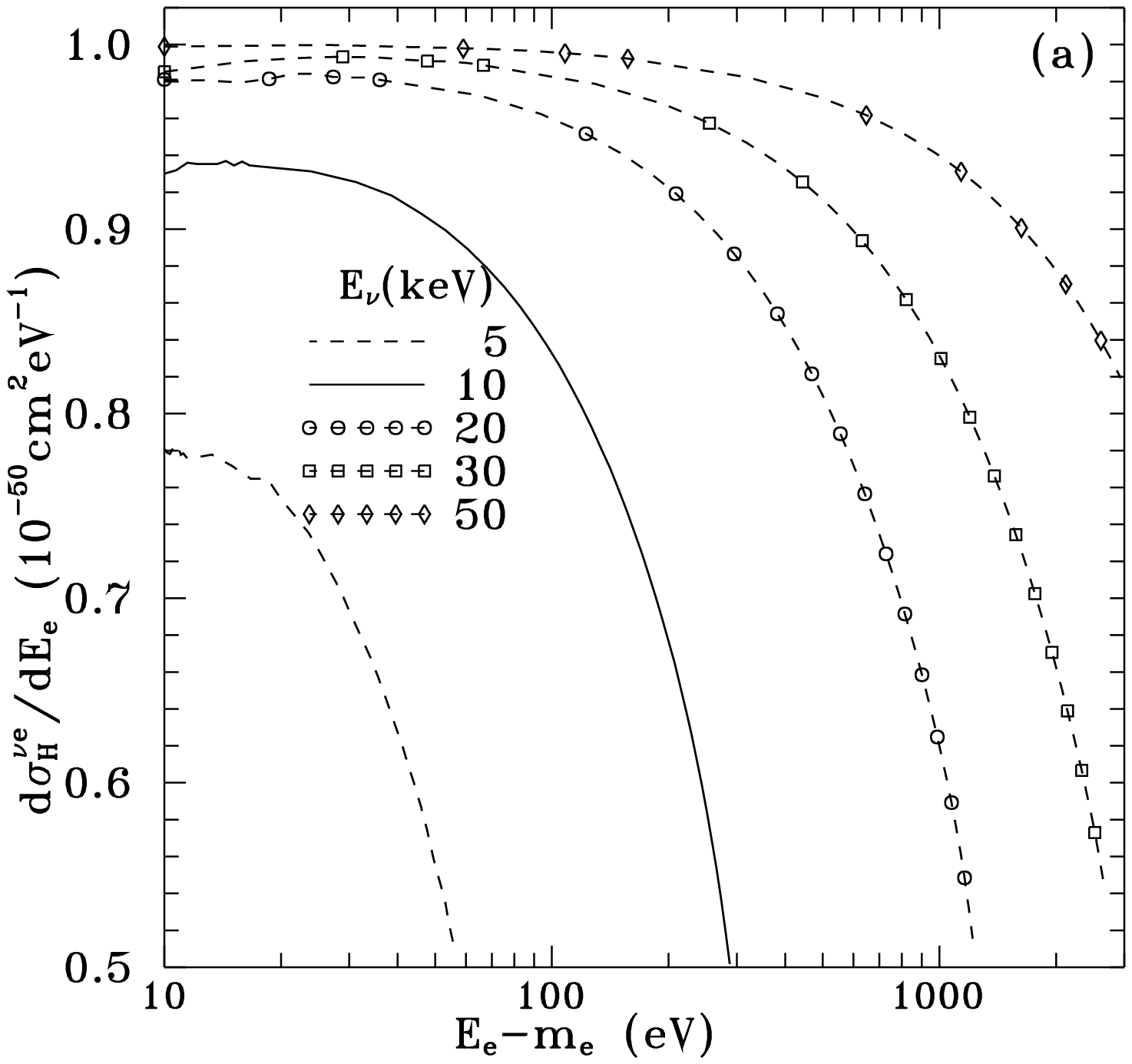,height=7.5cm, width=7.5cm}
\hspace{1.cm}\epsfig{file=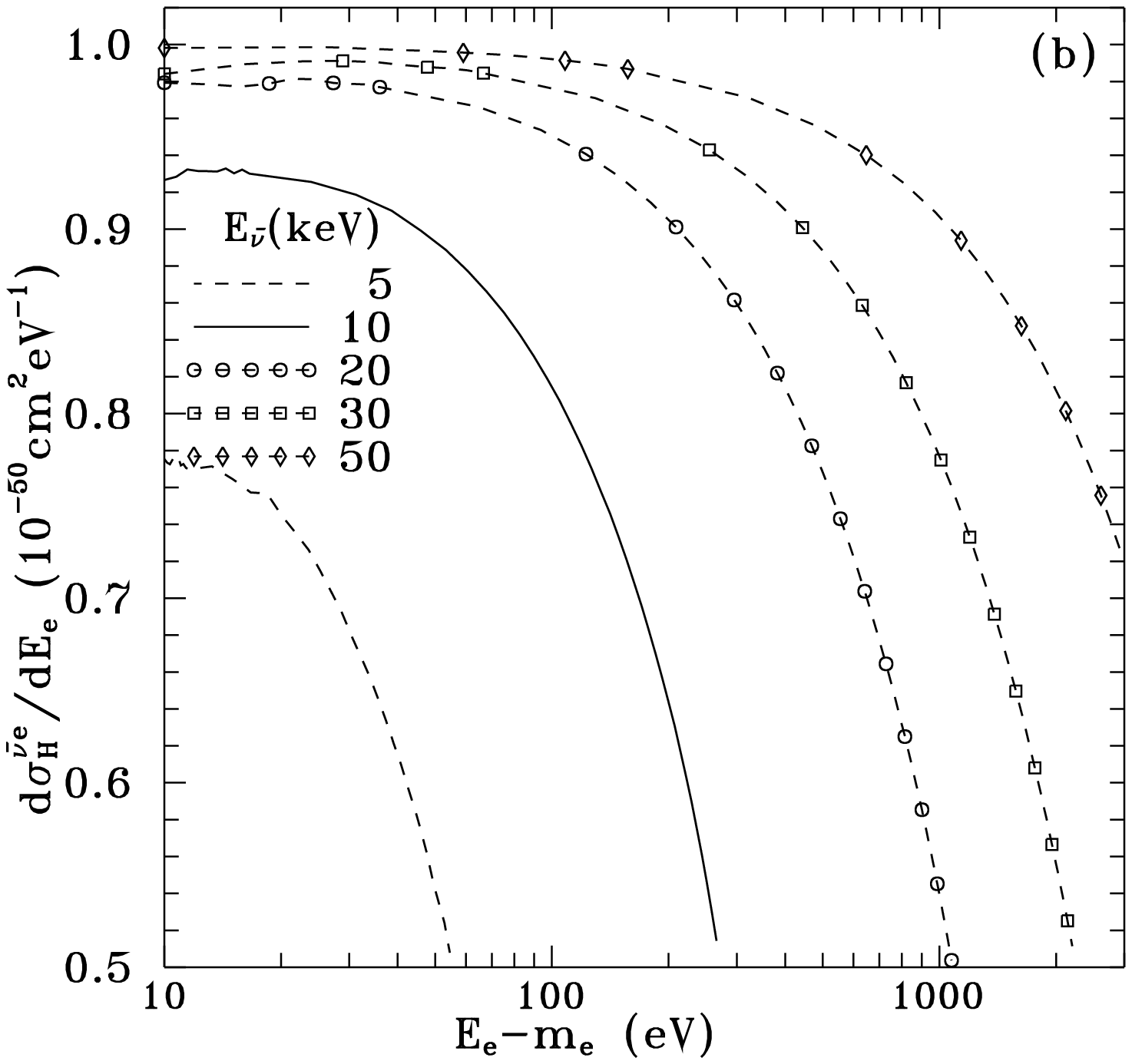,height=7.5cm, width=7.5cm}
\]
\[
\hspace{-0.5cm}\epsfig{file=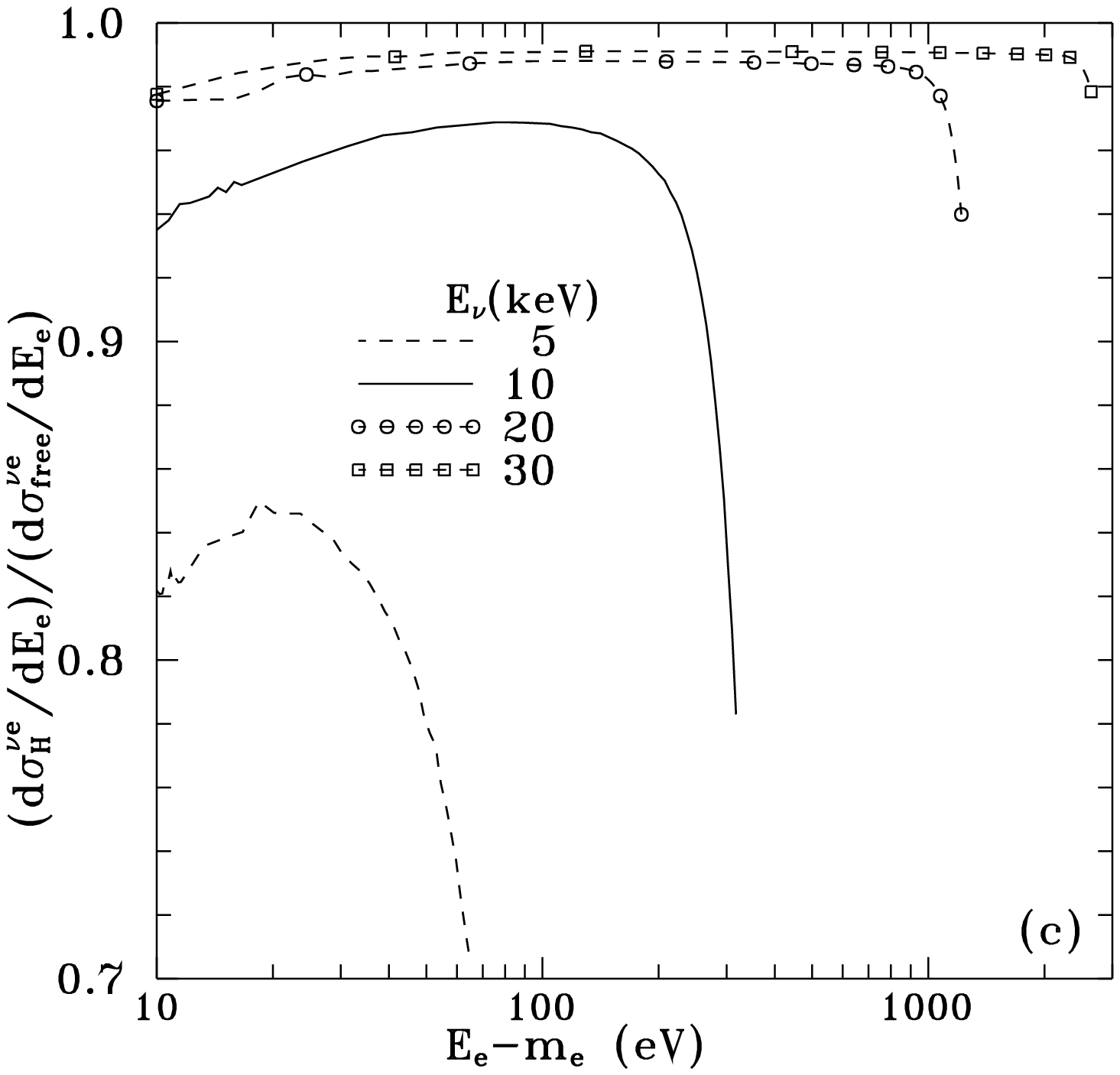,height=7.5cm, width=7.5cm}
\hspace{1.cm}\epsfig{file=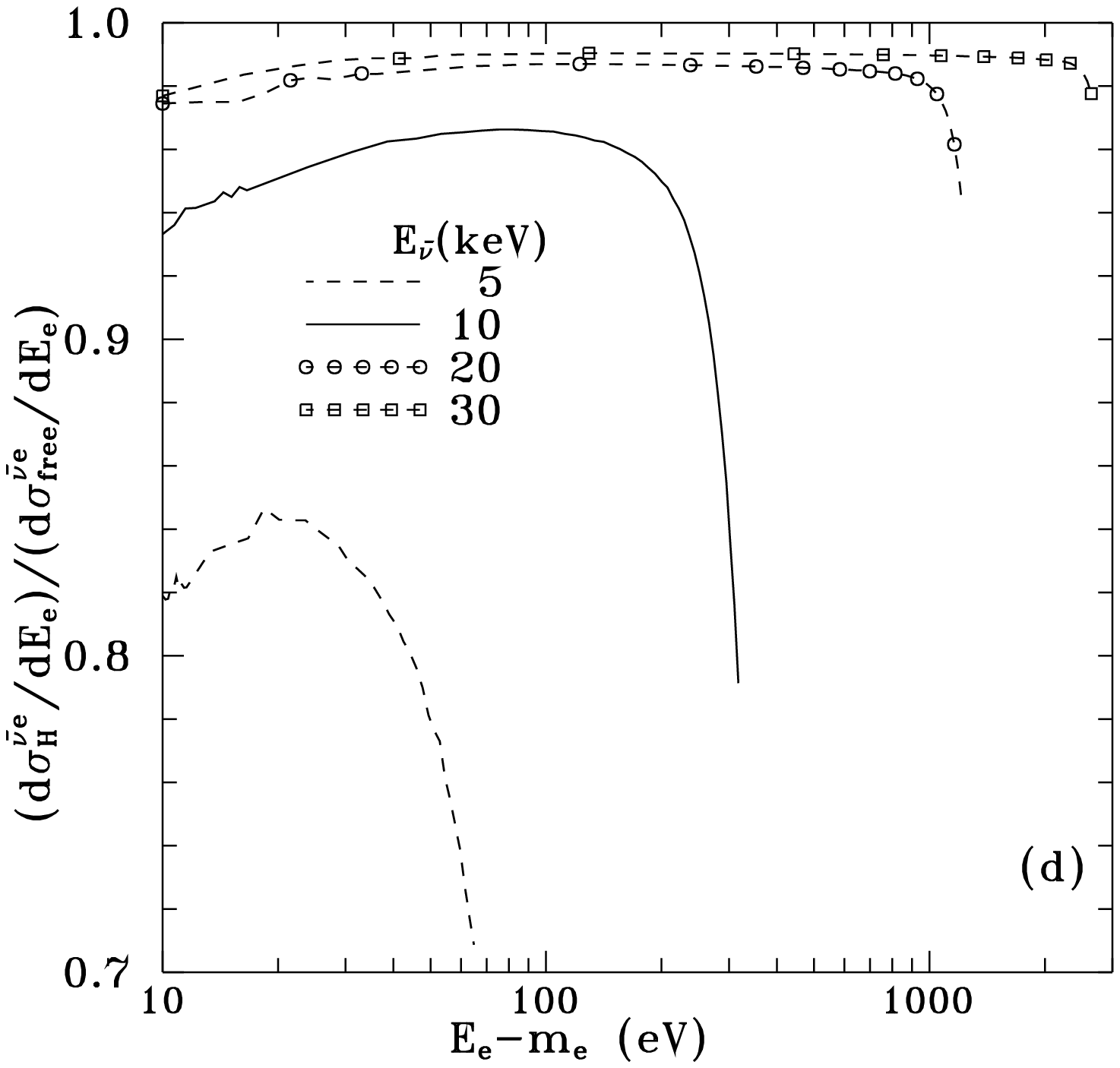,height=7.5cm, width=7.5cm}
\]
\caption[1]{The  electron energy distribution in   $\nu_e$ (a) or
$\bar \nu_e$ (b) ionization of $H$, and their respective ratio (c)
and (d)  to the corresponding  distributions when the initial
electron is assumed as free. Variables defined as in caption of
Fig.\ref{Free-energy-fig}.}
\label{H-energy-fig}
\end{figure}

The corresponding distributions for He and  Ne  atoms are shown  in
Figs.\ref{He-energy-fig} and  \ref{Ne-energy-fig}
respectively, using the wave functions of  Appendix A.2
and  the parameters tabulated in \cite{wave-functions}.
 For the  electron binding energies in  the various atoms
 (including those of Xe we discuss below),
we use the experimental values   \cite{e-binding}
\bqa
{\rm  ~~ He: }&&~~~  \epsilon^{He}_{1s}=-24.6 eV ~~~, \\
{\rm  ~~ Ne: }&& ~~~  \epsilon^{Ne}_{1s}=-870.2 eV ~~~, ~~~
\epsilon^{Ne}_{2s}= -48.5 eV ~~~, \epsilon^{Ne}_{2p}=-21.7 eV \\
{\rm  ~~ Xe: }&& ~~~  \epsilon^{Xe}_{1s}= -34561 eV ~~~,~~~
\epsilon^{Xe}_{2s}=-5453 eV~~~,~~~ \epsilon^{Xe}_{2p}=-4893 eV ~~~, \nonumber \\
&& ~~~ \epsilon^{Xe}_{3s}=- 1149 eV ~~~,~~~
\epsilon^{Xe}_{3p}=-961 eV ~~~,~~~ \epsilon^{Xe}_{3d}=-681.4 eV ~~~, \nonumber \\
&& ~~~ \epsilon^{Xe}_{4s}=-213.2  eV ~~~,~~~
\epsilon^{Xe}_{4p}=-145.7  eV ~~~,~~~ \epsilon^{Xe}_{4d}=-68.3 eV ~~~, \nonumber \\
&& ~~~ \epsilon^{Xe}_{5s}=-23.3  eV ~~~,~~~
\epsilon^{Xe}_{5p}=-12.5  eV ~~~.~~~
\eqa
As seen from  Figs.\ref{Free-energy-fig}-\ref{Ne-energy-fig}
the energy distribution in all $\nu_e (\bar \nu_e)$-induced
cross sections is maximal at small $E_e=E_4$.

Concerning  Figs.\ref{H-energy-fig}, \ref{He-energy-fig},
\ref{Ne-energy-fig}  we may  remark that
for $E_1$ in the few 10keV range, the  differential $\d \sigma/\d E_4$
ionization cross sections per electron are always at the level of
$10^{-50} cm^2 eV^{-1}$. Moreover, they  are   always
smaller than the "free electron" ones, approaching them  from bellow as
$E_1$ increases; see the (c) and (d) parts of each of these figures.
For a fixed $E_1$-value though, $\d \sigma/\d E_4$
always decreases as we go from Hydrogen to Ne and then to  Xenon.

\begin{figure}[p]
\vspace*{-2.cm}
\[
\hspace{-0.5cm}\epsfig{file=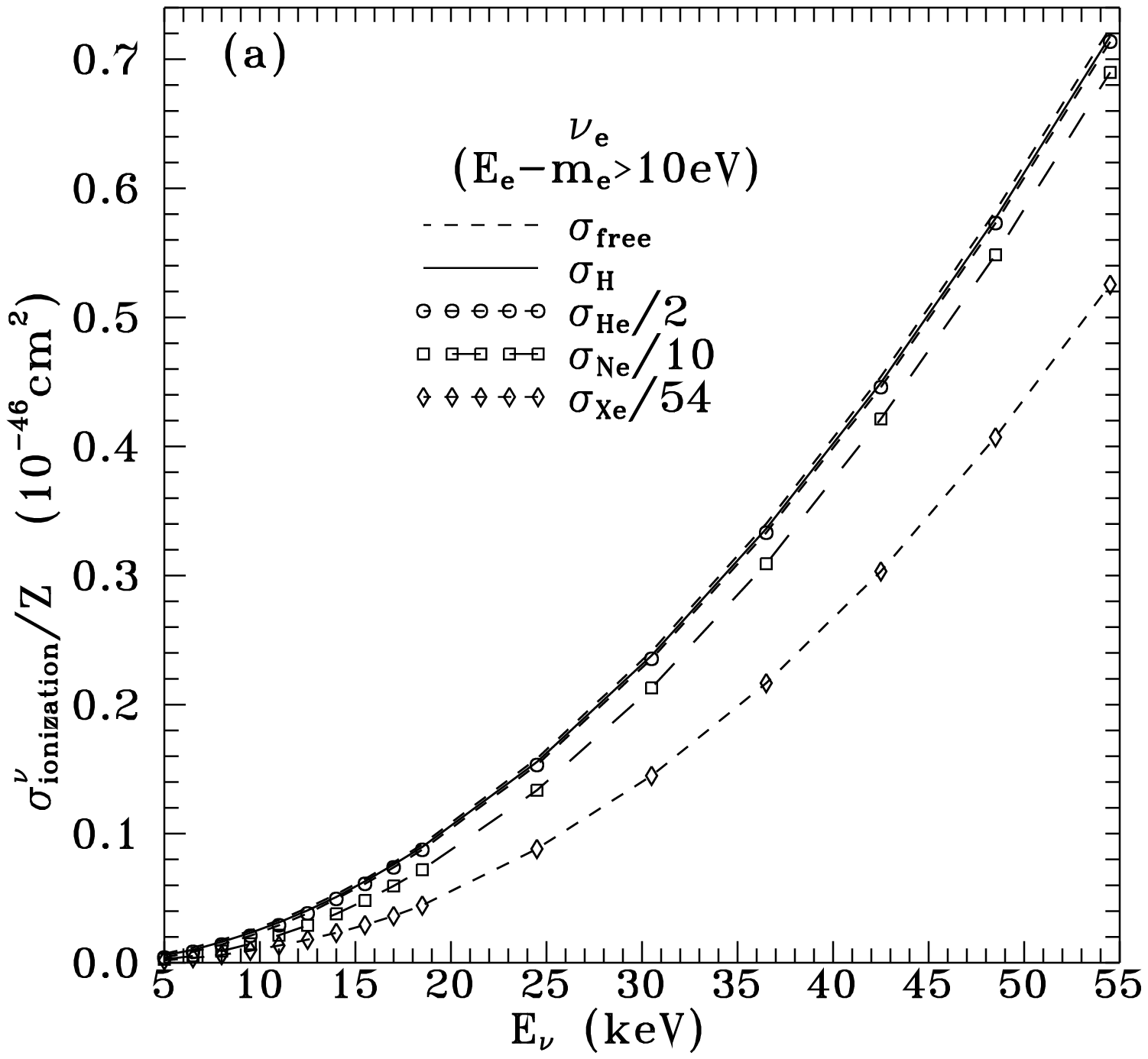,height=7.5cm, width=7.5cm}
\hspace{1.cm}\epsfig{file=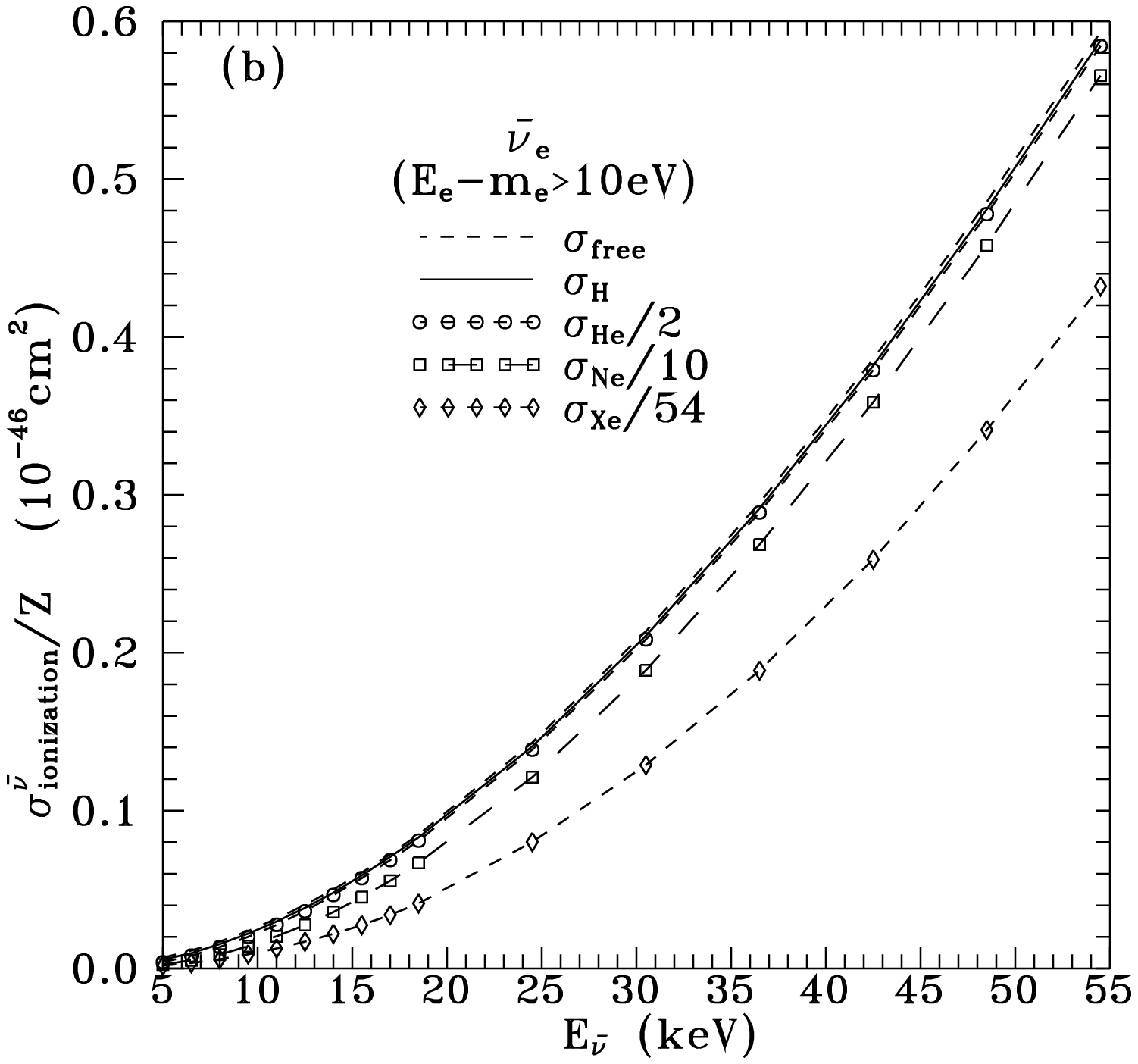,height=7.5cm, width=7.5cm}
\]
\[
\hspace{-0.5cm}\epsfig{file=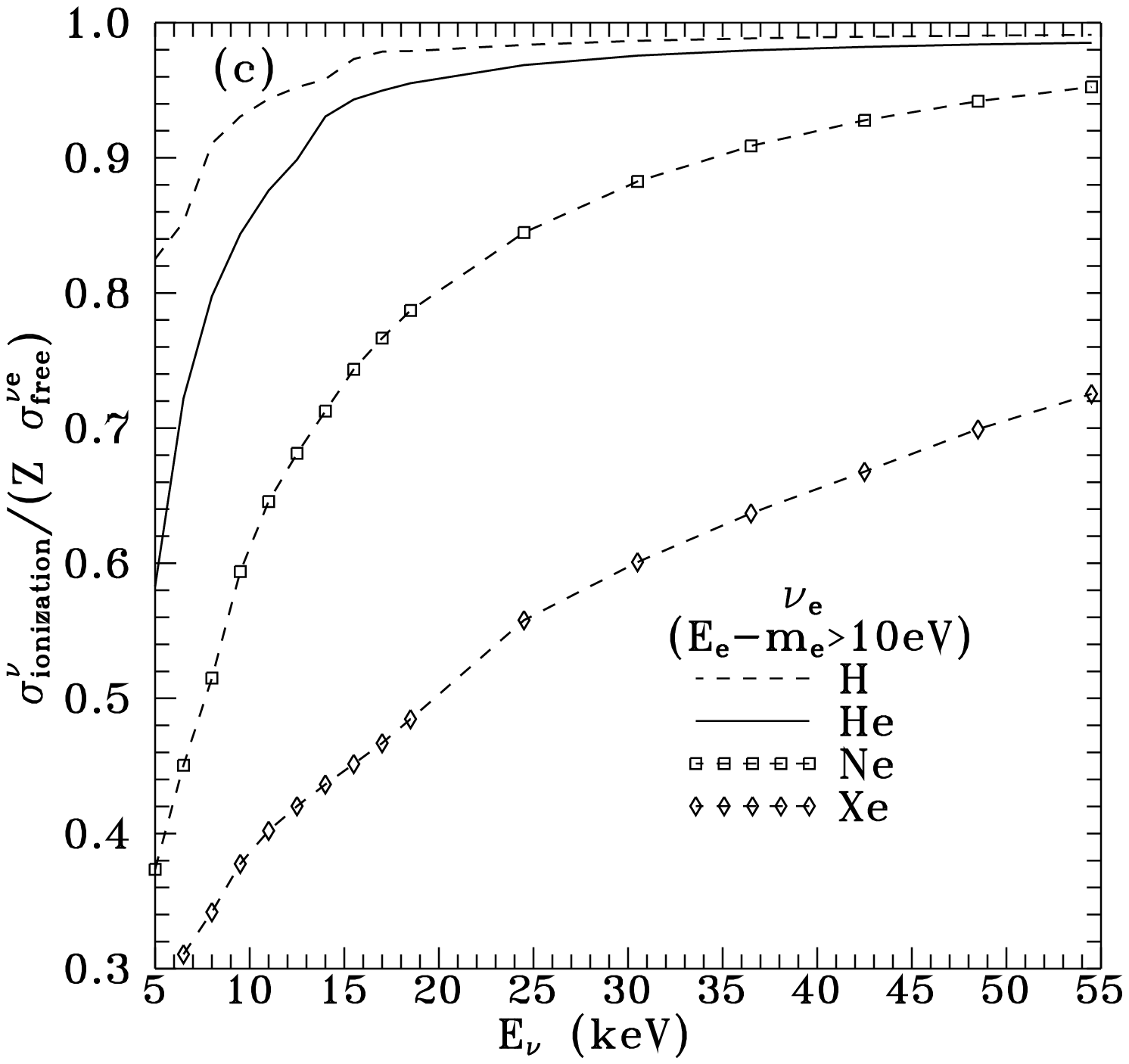,height=7.5cm, width=7.5cm}
\hspace{1.cm}\epsfig{file=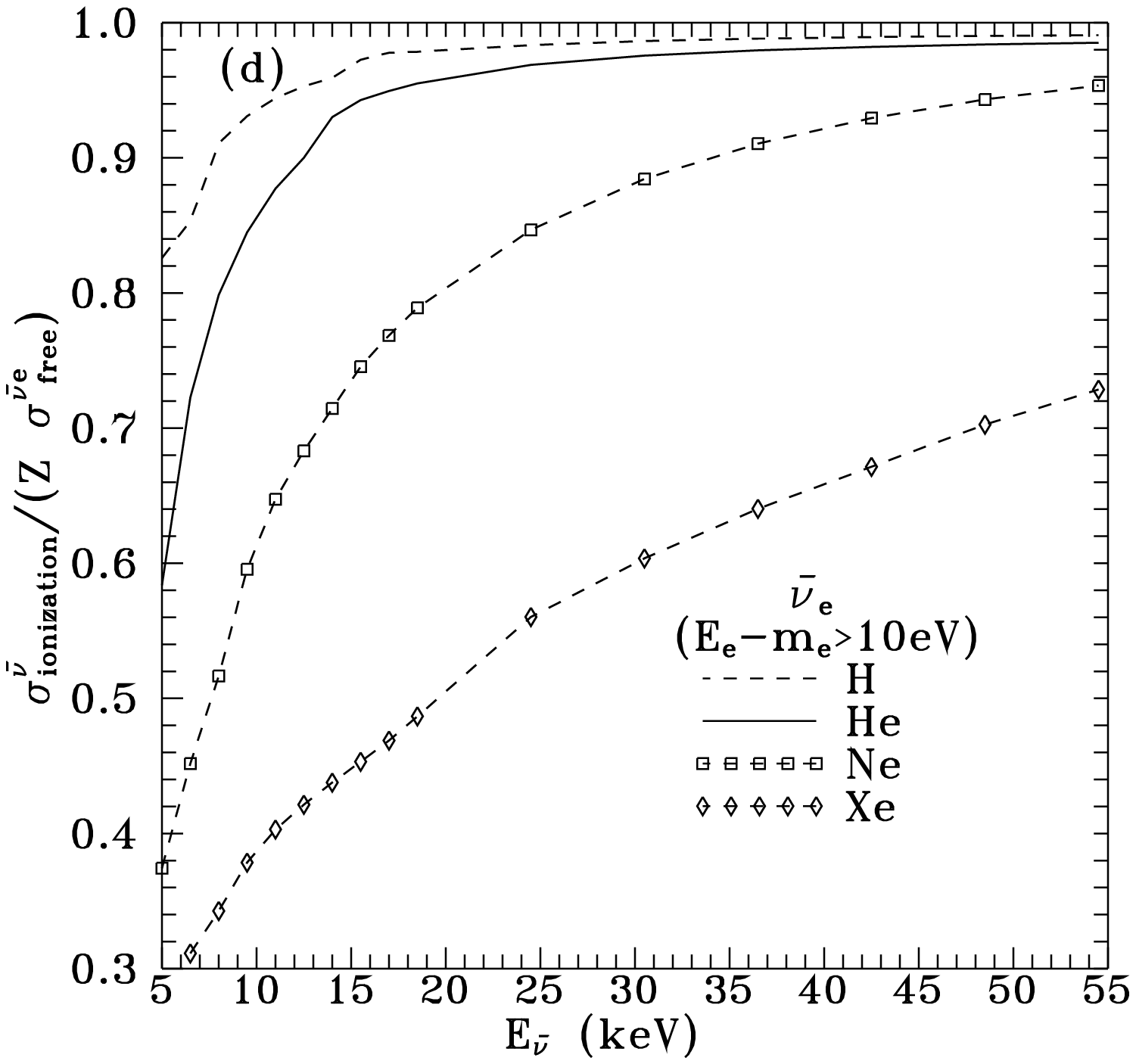,height=7.5cm, width=7.5cm}
\]
\caption[1]{The $\nu_e $ (a) [$\bar \nu_e$ (b)]
ionization integrated cross sections for H, He, Ne and Xe atoms divided by Z,
and the cross section off free electrons at rest, as a function
of the $\nu_e $  [$\bar \nu_e$ ] energy. The ratios of these ionization cross
section, to the free electron ones are given in (c) and (d) respectively.}
\label{sigma-fig}
\end{figure}

The integrated  $\nu_e(\bar \nu_e)$-ionization cross sections
for  final electron energies   $E_4>m_ e +10eV$,
are shown    in Fig.\ref{sigma-fig}. Here again the cross sections
are normalized to one electron per unit volume. In the same figure the
corresponding integrated cross section off free electrons are also included, as well
as our predictions for  Xe.
As seen from Fig.\ref{sigma-fig}a,b,  all  cross sections
are of the order of $ 10^{-47} cm^2$ at $E_1\sim 15 keV$,  increasing
with the neutrino energy. The ratios  of the integrated ionization cross sections to
the free electron one are shown  in Figs.\ref{sigma-fig}c,d. As expected,
the integrated ionization cross section is also always smaller than the free one,
 approaching it from below. For H and He the approach is very fast, but
it becomes much slower for heavier atoms; see  \eg  the results for
Ne and Xe. Thus at $E_1\sim 50 keV$ the Ne cross section is still
about $5\%$ smaller than the free $e$-one, while for Xe the decrease is
at the $\sim 30\%$ level.

Such integrated ionization cross sections for H, He and Ne have already
appeared in \cite{GPP}. The present results are consistent with
those. We should point out though that in \cite{GPP}, the integrated
cross sections were  calculated over the entire physical region $E_4 > m_e$.
Moreover, in that work we had used a very rough approximation for the He
wave function based on the $Z_{\rm eff}$-idea,
while the Ne results were based on the old fit of Tubis \cite{Tubis}.
Thus although the present results are consistent with those of  \cite{GPP},
prospective users should  rely more on  the present ones
based on  \cite{wave-functions}.
After all, it is only here that the energy distribution of the ionization
electron appears.\\

We have already estimated that for $\nu_e$ or $\bar \nu_e$ in the 10 keV level,
the oscillation length is about 150 meters; see (\ref{solar-oscillation}).
Consequently, the other neutrino flavors should also be generated, as
the initial $\nu_e$ or $\bar \nu_e$ proceeds in  a volume filed
with  a noble gas. Thus,  in a Tritium decay experiment,
$\bar \nu_\mu$ and $\bar \nu_\tau$ should   appear in relative amounts
determined by the mixing angles and the distance from the source.

\begin{figure}[p]
\vspace*{-2cm}
\[
\hspace{-0.5cm}\epsfig{file=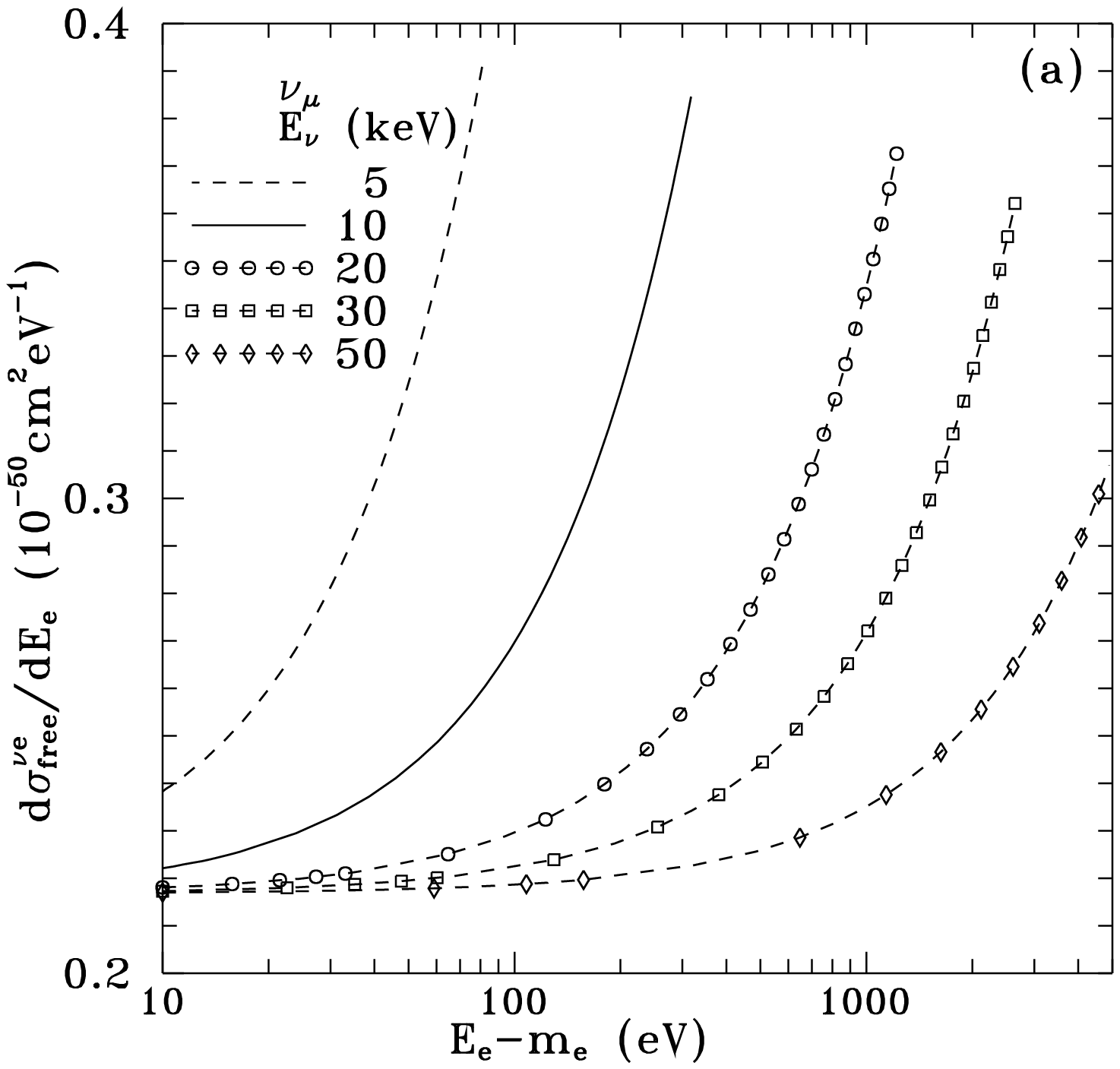,height=7.5cm, width=7.5cm}
\hspace{1.cm}\epsfig{file=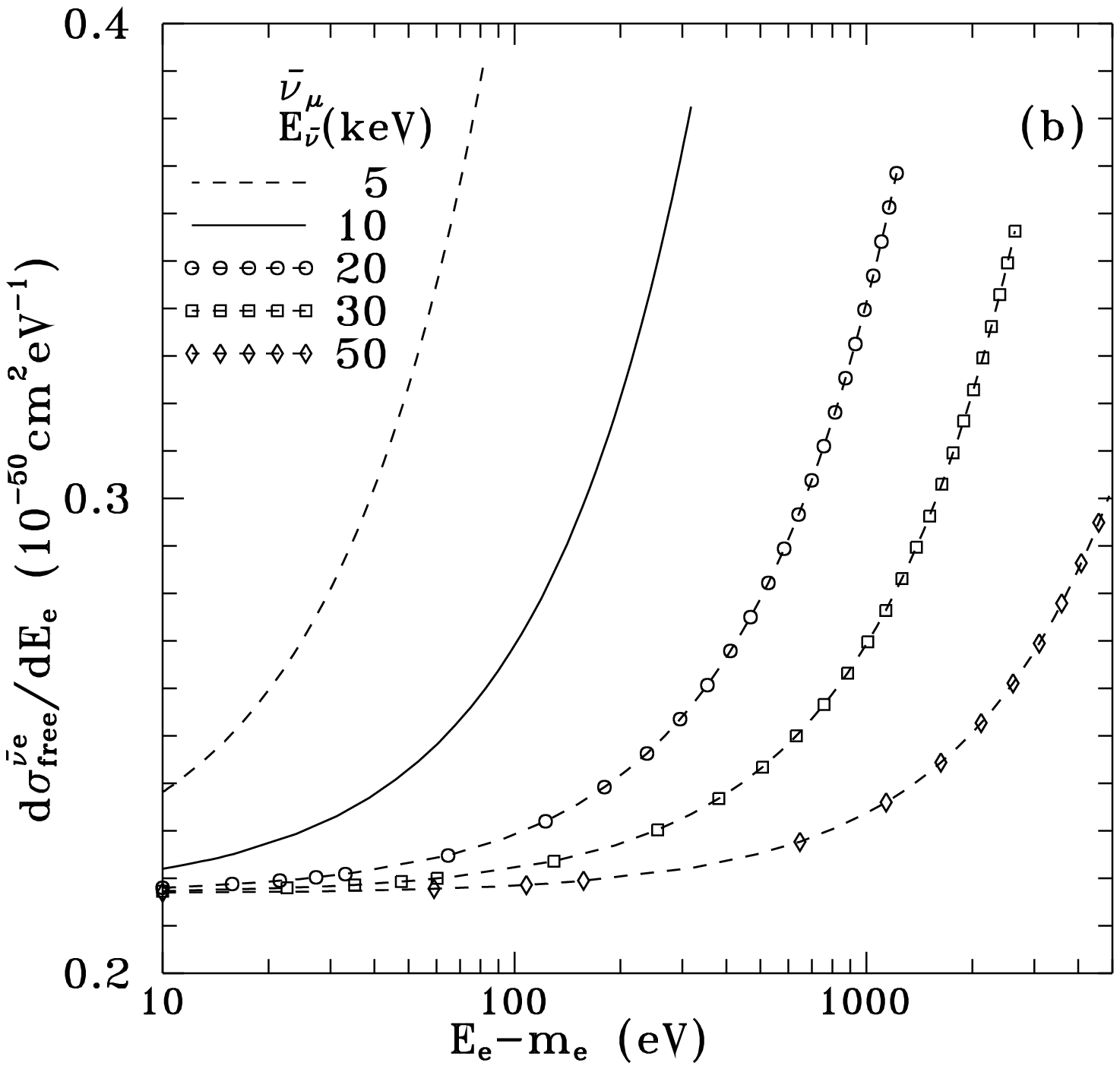,height=7.5cm, width=7.5cm}
\]
\caption[1]{The  energy distribution of the final electron in
$\nu_\mu e^- $ (a) or $\bar \nu_\mu e^- $ (b) scattering, at the rest
frame of a free initial $e^-$. Here $E_e\equiv E_4$ is the final
electron energy, and $E_\nu$ ($E_{\bar \nu}$) denote the incoming
neutrino (antineutrino) energy $E_1$. Identical results for $\nu_\tau$.}
\label{Free-energy-mu-fig}
\end{figure}

\begin{figure}[p]
\vspace*{-2cm}
\[
\hspace{-0.5cm}\epsfig{file=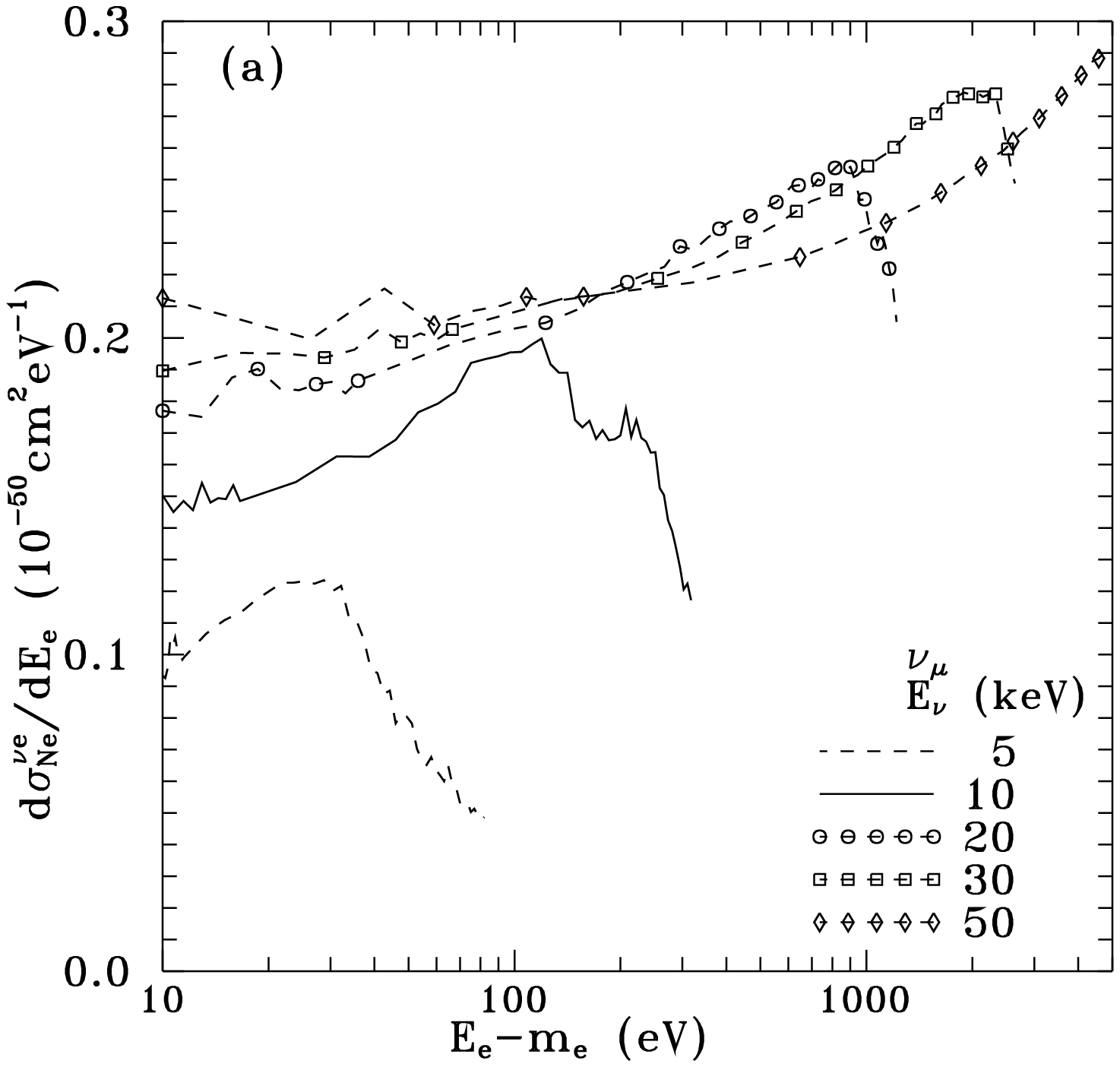,height=7.5cm, width=7.5cm}
\hspace{1.cm}\epsfig{file=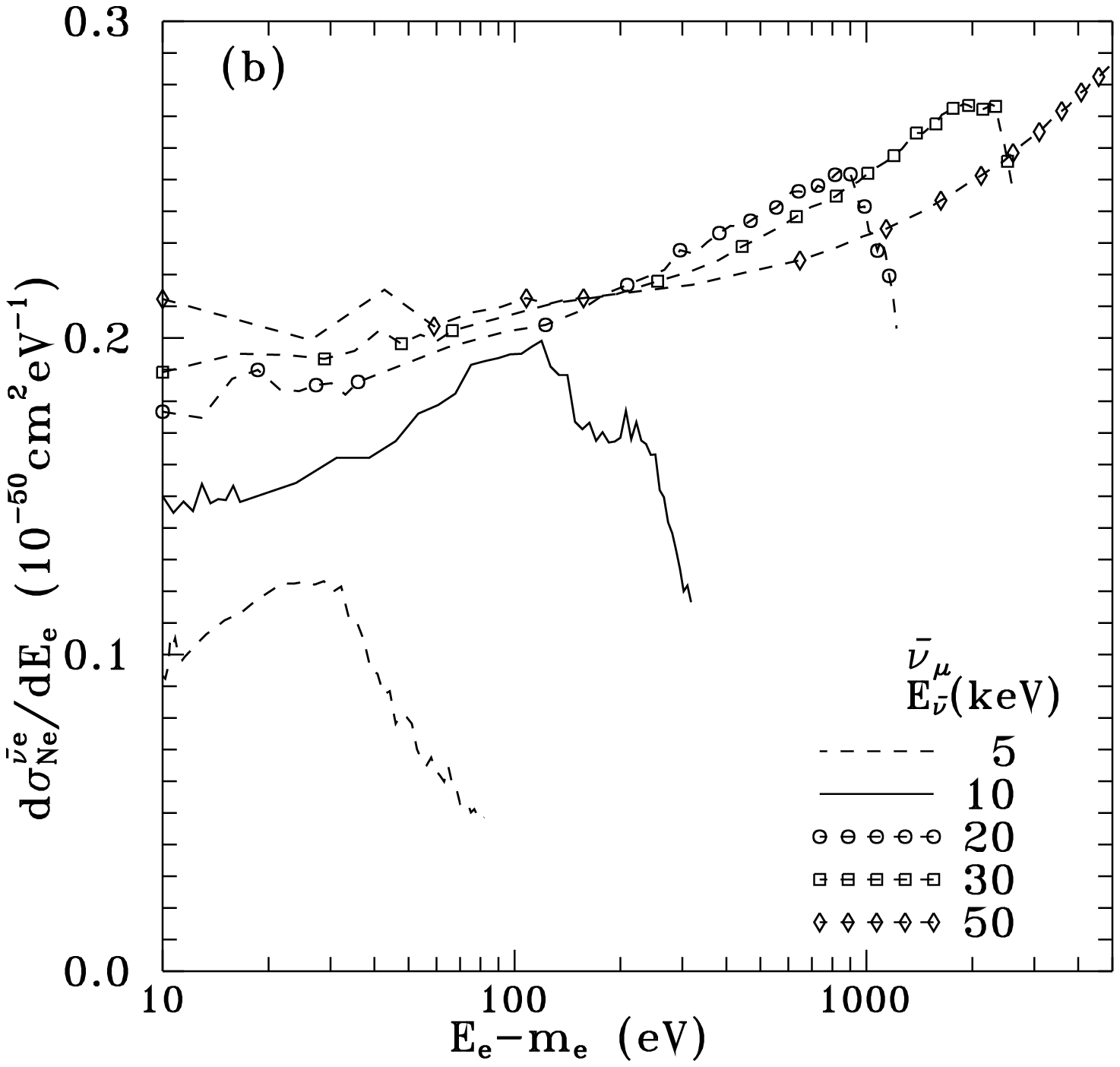,height=7.5cm, width=7.5cm}
\]
\[
\hspace{-0.5cm}\epsfig{file=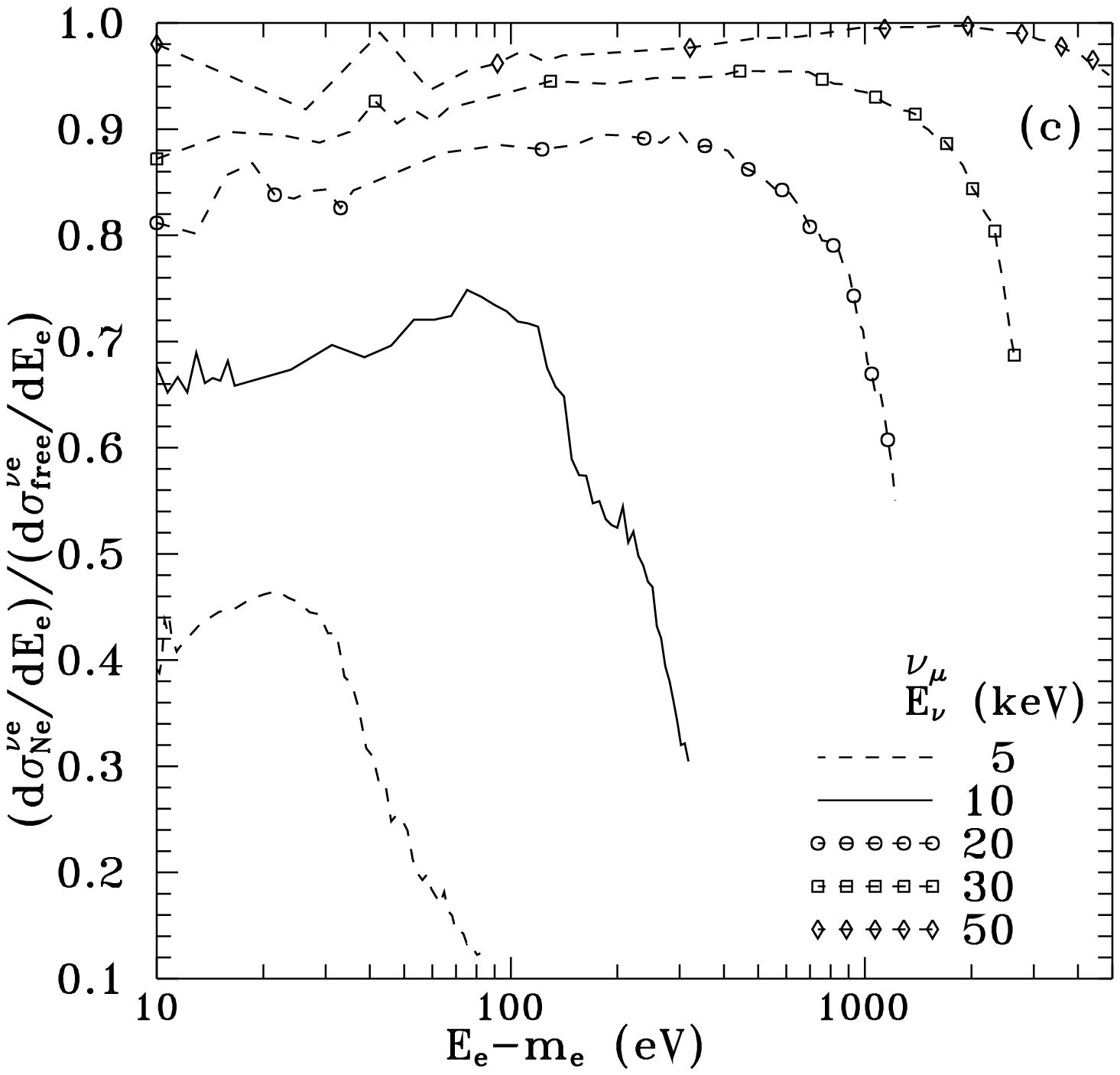,height=7.5cm, width=7.5cm}
\hspace{1.cm}\epsfig{file=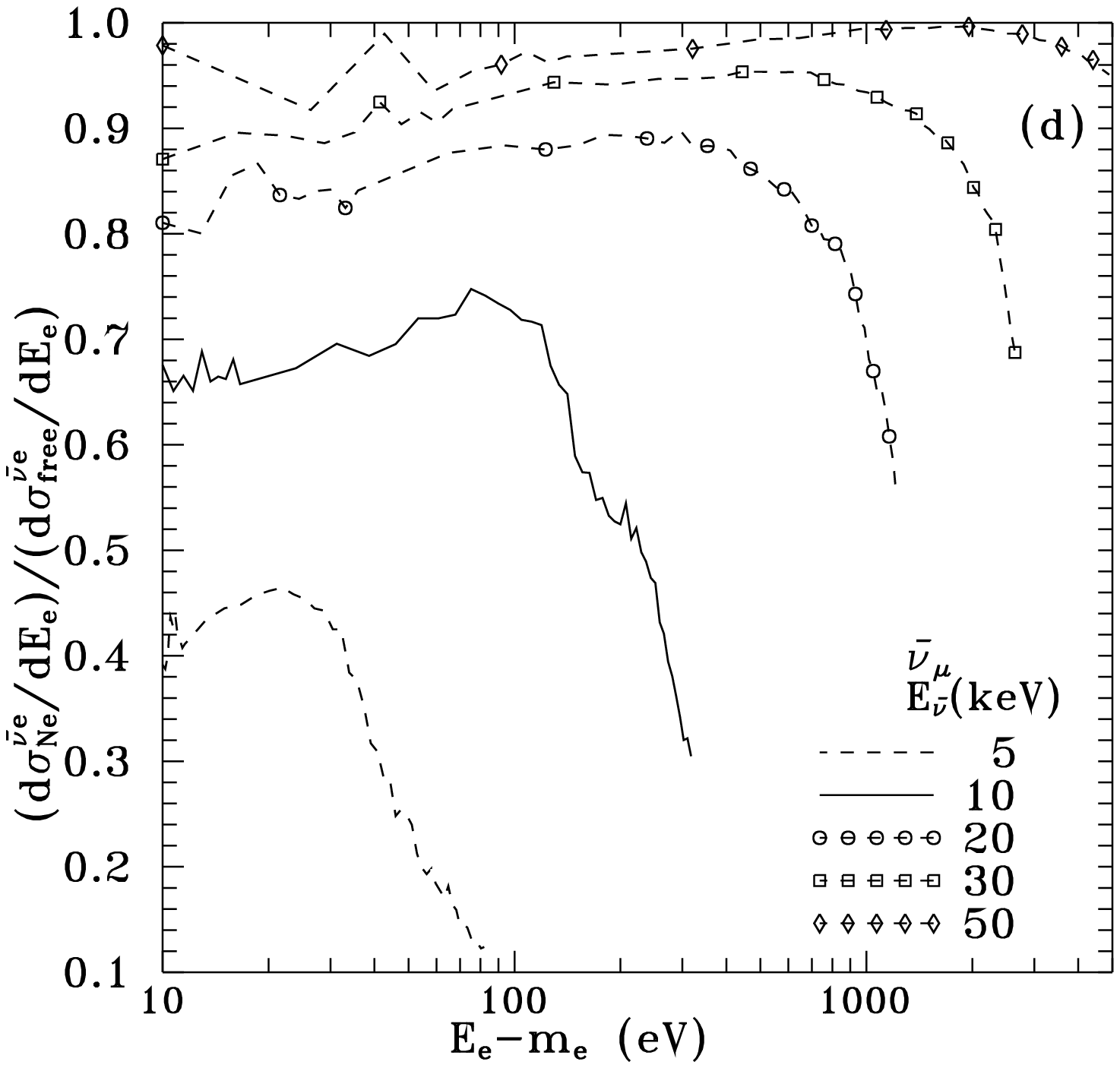,height=7.5cm, width=7.5cm}
\]
\caption[1]{The  electron energy distribution
in   $\nu_\mu$ (a) or $\bar \nu_\mu $ (b) ionization of $Ne$
normalized to one $e^-$ per unit volume, and their
respective ratio (c) and (d)  to
the corresponding  distributions when the initial electron is assumed
as free; see Fig.\ref{Free-energy-mu-fig}.
Identical results for $\nu_\tau$}
\label{Ne-energy-mu-fig}
\end{figure}

The treatment of the $\nu_\mu $ effects is exactly the same as for
$\nu_e$, the only difference being that in (\ref{Fsquared-free},
\ref{Fsquared}) we now have to use the second set of the neutrino
vector and axial couplings listed in (\ref{neutrino-couplings});
compare (\ref{Lagrangian}). The resulting differential cross section for
scattering off a free electron is then given
Fig.\ref{Free-energy-mu-fig}, which is strikingly different from
the corresponding $\nu_e(\bar \nu_e)$ result shown in
Fig.\ref{Free-energy-fig}. A similar situation arises
also for Ne  ionization; compare the $\nu_\mu$ induced ionization
shown in Fig.\ref{Ne-energy-mu-fig},
 to  the corresponding $\nu_e $ effect in Fig.\ref{Ne-energy-fig}.
 Again, the $E_e$
distributions tend to have  a local minimum at low final electron
energies in Figs.\ref{Free-energy-mu-fig},\ref{Ne-energy-mu-fig},
in contrast to the local maximum if
Figs.\ref{Free-energy-fig}-\ref{Ne-energy-fig}.

We have moreover found that for neutrinos (antineutrinos) at  the 10-20 keV range,
the respective ratio of
their integrated cross sections\footnote{Integrated for $E_e-m_e>10 eV$.}
  $\nu_\mu/ \nu_e$   ($\bar \nu_\mu/ \bar \nu_e$),
remains almost constant and equal to 0.42 (0.44), as we move from  the
  free electron case, to the Ne and  the Xe ionization ones.
  This suggests   that all  binding effects  in  the  individual
  $\nu_e$ and $\nu_\mu$ integrated cross sections, are   cancelling  in their ratio;
  for which the simple expression (\ref{sigma-nu-free}) is   adequate.

It is worth remarking also on the basis of Figs.\ref{Free-energy-mu-fig}
and \ref{Ne-energy-mu-fig}, that the $\nu_\mu$ and $\bar \nu_\mu$
results are almost identical. This is due to the fact that
$\tilde v_e \simeq 0$,  for the value of the Weinberg angle we use
(compare (\ref{neutrino-couplings})),
which makes the squared amplitudes in (\ref{Fsquared-free}, \ref{Fsquared})
almost $ s \leftrightarrow u$ symmetric. It seems therefore that
a Tritium experiment may also give  information on
the value of the Weinberg angle at the keV scale \cite{Giomataris1}.
The $\nu_\tau$ results are of course  identical to those
for $\nu_\mu$; compare (\ref{Lagrangian}).\\

We next briefly address the problem of the flavor
oscillation of the neutrino induced noble gas ionization.
Since for reasonable gas-densities the vacuum oscillation
treatment should be adequate, the oscillation probabilities may be written as
\bqa
P(\nu_e\to \nu_\mu+\nu_\tau )&=& \sin^2 (2 \theta_{12} ) \cos^4(\theta_{13})
\sin^2 \Big ( \frac{\Delta_{12} L}{4 E_1} \Big ) +
\sin^2 (2 \theta_{13} ) \sin^2 \Big ( \frac{\Delta_{13} L}{4 E_1} \Big ) ~,
\label{oscillation-prop1} \\
 P(\nu_e \to \nu_e) & = & 1- P(\nu_e\to \nu_\mu+\nu_\tau ) ~~,
\label{oscillation-prop2}
\eqa
assuming   three active   and no  sterile neutrinos. Here
$\theta_{12}, ~\theta_{23},~\theta_{13}$  are
the 3 mixing angles, while   $\Delta_{ij}\equiv m_j^2-m_i^2$
denote   the mass differences between the neutrino masses satisfying
$|\Delta_{13}|\simeq |\Delta_{23}| \gg |\Delta_{12}|$  \cite{Reines, neutrino-data},
and $L$ is the distance from the source.
We note that  there is no dependence on the neutrino CP-violating phase $\delta$
in  (\ref{oscillation-prop1}, \ref{oscillation-prop2}),
and that these same formulae
 describe antineutrino oscillations  also.

Thus, as an initially produced $\nu_e$  transverses \eg a
Ne target, the  ionization cross section per electron varies
with $L$ as
\bq
 \frac{ \d \sigma_{\rm Ne}}{\d E_4}= [1- P(\nu_e\to \nu_\mu+\nu_\tau )]
\cdot \frac{ \d \sigma_{\rm Ne}^{\nu e}}{\d E_4}\Bigg |_{\nu=\nu_e}
+P(\nu_e\to \nu_\mu+\nu_\tau )
\cdot \frac{ \d \sigma_{\rm Ne}^{\nu e}}{\d E_4}\Bigg |_{\nu=\nu_\mu}
~~ , \label{dsigma-dE4-oscillation}
\eq
 in the r.h.s of which (\ref{dsigma-dE4-H}, \ref{He-Xe-atoms})
should be used for $\nu=\nu_e$ and $\nu=\nu_\mu$ respectively.
Correspondingly for antineutrinos.

Using the   experimental  $\Delta_{ij}$ values,
  we find that the oscillation length of the first term
of $P(\nu_e\to \nu_\mu+\nu_\tau )$ (see  (\ref{oscillation-prop1}))
is about 150m for an  incoming neutrino energy of $E_1\simeq 10 keV$,
while the oscillation length of the second  term  is only  about 5.6m.
Since $\theta_{13}$ is known to be very small, the picture created by
(\ref{dsigma-dE4-oscillation}) will then consist of a
few hundred meter oscillation, modulated by a much weaker one of a few meter
size \cite{neutrino-data}. The strength of the modulation is only
determined by $\theta_{13}$, and  the ratio of the
$\nu_\mu$ to $\nu_e$ cross sections. As we have  already stated,
an enhancement of the theoretical accuracy of this ratio immediately implies also
an increase to  the $\theta_{13}$ sensitivity.

Depending, therefore, on the achievable experimental accuracy,
the  study $\bar \nu_e$ oscillations
in Tritium decay may help further constraining
the neutrino mixing angles and masses. This would be  most interesting
for $\theta_{13}$, for which information might be derived if  a future
experiment  manages to be sensitive to both oscillation
lengths  governing  (\ref{oscillation-prop1}).  \\

\section{Conclusions}

In this paper we have presented a  formalism for the ionization
of atoms by bombarding them with neutrinos of any flavor in the
keV energy range. The interest in this energy range originates from
the fact that the  oscillations lengths for neutrinos in the few keV
range become rather small, allowing the
 possibility of  studying the oscillations observed
in the solar neutrino and KAMLAND experiments, by means of a terrestrial
experiment of  small size. A highlight of such an experiment is to
improve the constraint on $\theta_{13}$.  Neutrinos in this
energy may be obtained from various possible beta decays;
most notably Tritium decay producing antineutrinos.

Motivated by this, we have undertaken
the present   extensive study of the ionization of H and the noble atoms by neutrinos.
To this purpose, we have developed a method based on spin independent atomic
wave functions, for which we use the very accurate
Roothann-Hartee-Fock (RHF) wave functions  listed in \cite{wave-functions}.
 The energy of the bound electron in each atomic state is fixed by its binding energy,
while  its momentum varies according to the distribution determined by the
momentum wave function, thereby generally forcing the electron to  get off-shell.
 The method is very accurate for treating H  and the most interesting noble gases
He, Ne and   Ar.

On the other hand, it is less accurate for  Xe, where considerable
spin dependence appears.  Thus, for Xe we have simply given the results
of the present method for the integrated ionization cross section, whose
accuracy should be lying at  the 10\% level or so.  This completes the
overall view of the neutrino ionization of the noble gases.
If the use of Xe in such an experiment is finally decided though, than of course
a special treatment would be needed, based on spin depended atomic wave functions.\\

We next  proceed
to summarize our extensive results.
We have found that at the 10--20 keV neutrino energy range,
the differential cross sections
$\d \sigma /\d E_e$   are  at the $10^{-50} cm^2/eV$ level
for $\nu_e (\bar \nu_e)$-induced processes, while the integrated cross section
is of the order of $10^{-47} cm^2$.
At the  10--20 keV neutrino energy range,  atomic
effects are very important and cannot be ignored. They reduce the Ne ionization
cross section per electron by almost 20\%,
while for Xe  the reduction reaches the  factor of two level.
Of course, as the incoming neutrino energy increases
beyond \eg the 100 keV region, all these "per-electron"
ionization cross sections approach the neutrino cross section off free
 electrons, always from below.

We have also compared the $\nu_e$ induced
reactions, with those induced by  $\nu_\mu $ or
 $\nu_\tau $; the later two being equal  to each other.
 The difference between the two comes from the fact that
 $\nu_e$ reactions involve both charged and neutral current diagrams,
 while for $\nu_\mu, ~\nu_\tau $ only neutral currents contribute.
As a result,  the  energy distribution of the final electron
  in the  $\nu_\mu$  reactions  tends to have
a local minimum at low electron  energies, in contrast to the
local maximum expected for the corresponding $\nu_e$ effect.
Moreover for 10--20 keV neutrinos or antineutrinos, the ratio
of their integrated  cross sections is found to be largely
independent of any atomic binding effects.

As an overall conclusion we  may state
that if such small cross sections  become measurable
one day,   neutrino  atomic ionization experiments
may be useful for testing  the electroweak theory at keV energies
and studying  the neutrino  mixing.

\vspace{1cm}
\noindent{\large\bf{Acknowledgement}}\\
\noindent
The support of the
``Bundesministerium f\"ur Bildung, Wissenschaft, Forschung und
Technologie'', Bonn under contract 05HT1PEA9,
and the support by European Union under the RTN contracts
HPRN-CT-2000-00148 and
MRTN-CT-2004-503369, are  gratefully acknowledged. P.I.P.  is also grateful
to the  Institut f\"{u}r Physik, Universit\"{a}t Dortmund,
for the hospitality extended to him during part  of this work.

\newpage
\appendix


\renewcommand{\theequation}{A.\arabic{equation}}
\renewcommand{\thesection}{Appendix A.\arabic{section}}
\setcounter{equation}{0}
\setcounter{section}{0}

\section{Kinematics}

In the rest frame of the atom, defining the $\hat z$-axis along
the direction of the incoming neutrino, and the $xz$-plane as the
plane where  the (final) ionization electron  lies, we   write,
(compare  (\ref{process}))
\bq
P_1^\mu= \left( \begin{array}{ccc}
E_1\\
0\\
0\\
E_1 \end{array} \right) ~,~
P_2^\mu= \left( \begin{array}{ccc}
E_2\\
p_2 \sin\theta_2 \cos\phi_2 \\
p_2 \sin\theta_2 \sin\phi_2 \\
p_2 \cos\theta_2  \end{array} \right) ~,~
P_4^\mu= \left( \begin{array}{ccc}
E_4\\
p_4 \sin \theta_4  \\
0 \\
p_4 \cos\theta_4  \end{array} \right) ~,~ \label{kinematics}
\eq
where $E_1$ is the energy of the initial neutrino;
$(E_2, p_2, \theta_2, \phi_2)$ are the energy,
momentum and  angles of the
electron bound inside  the atom; and $(E_4, p_4, \theta_4)$ are the energy,
momentum and polar angle of the freely moving final ionization electron.
By definition, the range of these angles is
\bq
0<\theta_4 <\pi~~, \label{theta4-range}
\eq
\bq
0<\theta_2 <\pi~~~,~~~ 0<\phi_2< 2\pi ~~~.  \label{theta2-phi2-range}
\eq

The standard variables defined in (\ref{Mandelstam})  become
\bqa
s&=&\tilde m^2+2 E_1 (E_2-p_2 \cos\theta_2)  ~~, \nonumber \\
t&=& m_e^2-2 E_1 (E_4-p_4\cos\theta_4) ~~, \nonumber \\
u&=& 2 E_1( E_4-p_4 \cos\theta_4-E_2 +p_2 \cos\theta_2) ~~.
\label{Mandelstam2}
\eqa
where the definition (\ref{mtilde2}) is used, and (\ref{Mandelstam1})
is of course satisfied.

The requirement of \bq P_3^2=(P_1+P_2-P_4)^2=0 ~~
\label{theta4-equation} \eq implied by the negligibly small
neutrino mass,  leads to an  equation determining $\cos
\theta_4$  in terms of  $E_4$ and $E_1$.  In the free electron
case, this  is linear in $\cos \theta_4$, and it can  be solved
immediately leading to  (\ref{dsigma-nu-free}).

For bound electrons, however,  (\ref{theta4-equation}) leads to a
quadratic equation in $\cos \theta_4$, which also depends on the
momentum and spherical angles of the bound electron. It is then
important to discriminate between the two mathematically possible
solutions, among which only one is physically acceptable. To do
this we first note that the general solution of
(\ref{theta4-equation}), may  be written as
\bq
\tan\Big
(\frac{\theta_4}{2} \Big )= \frac{\zeta_1 \pm
\sqrt{\zeta_1^2-\zeta_2}}{2\zeta_3}~~~,~~~ \label{theta4-solution}
\eq where \bqa
\zeta_1&=& 2 p_4 p_2 \sin\theta_2 \cos\phi_2 ~~, \nonumber \\
\zeta_2 &=& 4\Big \{ [E_4 (E_1+E_2)-\xi]^2-p_4^2 (E_1+p_2 \cos\theta _2)^2 \Big \}
~~, \nonumber \\
\zeta_3 &=& E_4 (E_1+E_2) -\xi +p_4 (E_1+p_2 \cos\theta_2) ~~,
\label{zeta-param}
\eqa
\bq \xi= -E_1 p_2 \cos\theta_2 +E_1 E_2 +\frac{m_e^2 + \tilde
m^2}{2} ~~. \label{xi-param} \eq A detail study indicates that
whenever both $\theta_4$  solutions of (\ref{theta4-solution})
satisfy  (\ref{theta4-range}), the physically acceptable one is
given by the upper (lower) sign of (\ref{theta4-solution}),
depending on whether\footnote{According to the first of
(\ref{zeta-param}), this is equivalent to $\cos\phi_2$ being
positive (negative) respectively.} $\zeta_1  >0$ ($\zeta_1  <0$)
respectively. This physically acceptable $\theta_4$  solution, is
by its very definition, a continuous function of
the ionization electron energy, and  the bound electron's momentum
and angles entering the integration in (\ref{dsigma-dE4-H}).

Once the physically acceptable  $\theta_4$ solution in
(\ref{theta4-solution}) has been identified,
$\d u/ \d E_4$ entering (\ref{dsigma-dE4-H})  is  determined from
\bqa
\frac{\d  u}{\d E_4}&= & -2 E_1 \Big (1-\frac{E_4}{p_4}\cos\theta_4
+p_4 \frac{\d \theta_4}{\d E_4}\sin\theta_4 \Big )~~,
\nonumber \\
\frac{\d \theta_4}{\d E_4}&=&
\frac{E_4 [E_1 \cos\theta_4 +p_2 (\cos\theta_2 \cos\theta_4+
\sin\theta_2 \sin\theta_4\cos\phi_2)]-(E_1+E_2)p_4}
{p_4^2 [E_1 \sin\theta_4 +p_2 (\cos\theta_2\sin\theta_4
-\sin\theta_2\cos\theta_4 \cos\phi_2)]}~. \label{du-dE4}
\eqa

Finally, the requirement $\zeta_1^2 \geq \zeta_2$ (note  the square root in
 (\ref{theta4-solution})) imply that
 the range of  $E_4$  is determined by
\bq
 E_4^{(2)}  \leq E_4 \leq E_4^{(1)} ~~, \label{E4-range-a}
\eq
with
\bqa
E_4^{(1,2)}&=&\frac{\xi (E_1+E_2)\pm \xi_1
\sqrt{\xi^2-m_e^2 [(E_1+E_2)^2-\xi_1^2]}}{(E_1+E_2)^2-\xi_1^2}~~,
\nonumber \\
\xi_1 &=& \sqrt{(E_1+p_2 \cos\theta_2)^2+p_2^2\sin^2\theta_2\cos^2\phi_2}~~.
\label{E4-range-b}
\eqa\\

\section{The Atomic Wave Functions}

The atomic  wave functions in the momentum space needed in (\ref{dsigma-dE4-H}),
 are related to the coordinate wave functions by
\bq
\wtil \Psi_{nlm}(\vec k)= \int \d^3r \Psi_{nlm}(\vec r) e^{-i
\vec k\cdot \vec r}~~, \label{Psitil}
\eq
where $nlm$ denote the
usual quantum numbers characterizing atomic states. The radial
momentum wave function
 defined by the first function  in the r.h.s of\footnote{$Y_{lm}(\hat k )$
 is the usual spherical harmonic function depending on the momentum angles.}
\bq
\wtil \Psi_{nlm}(\vec k)=(-i)^l \wtil R_{nl}(k) Y_{lm}(\hat k )
~~, \label{Rtil}
\eq
is  related to the radial wave function in coordinate space
through
\bq
\wtil R_{nl}(k)=4\pi \int_0^\infty \d r r^2 R_{nl}(r) j_l(kr)~, \label{Rtil-1}
\eq
where $j_l(kr)$ is a spherical Bessel function. The normalization is
such that
\bq
 \int \frac{\d^3 k }{(2\pi)^3} |\wtil \Psi_{nlm}(\vec k ) |^2=
 4 \pi \int_0^\infty \frac{ k^2 \d k }{(2\pi)^3} |\wtil R_{nl}( k ) |^2=1
~~ . \label{normalization}
\eq\\

We next turn to the explicit form of
momentum wave functions for the ground state of the various atoms.
For Hydrogen, the standard ground state wave function is
\begin{equation}
\wtil \Psi_{100}( k )= \frac{1}{\sqrt{4\pi}} \wtil R_{10}(k) =
\frac{8\sqrt{\pi}(Z_H)^{5/2}}
{(k^2+(Z_H)^2)^2}~~, \label{H-function}
\end{equation}
with $Z_H=m_e\alpha $ being the inverse Bohr radius of the H-atom.
In all cases involving   integrals over the Hydrogen momentum wave
function, we integrated over the range  $0\leq k \lsim 3 Z_H$,
which describes very accurately the relevant momentum range of the bound electron.\\

For the heavier atoms we follow \cite{wave-functions}, based on
the Roothann-Hartee-Fock (RHF) approach, also explained in
\cite{Slater}. Accordingly,  the radial momentum wave functions
are written as
\bq
\wtil R_{nl}(k)=\sum_j C_{jln} \wtil S_{jl} (k)
~~, \label{Rnl-RHF}
\eq
in terms of  the  RHF functions $\wtil S_{jl} (k)$ in momentum space,
related  to  $ S_{jl} (r)$  given
in ref. \cite{wave-functions}  through
\bq
\wtil S_{jl}(k)=4\pi
\int_0^\infty \d r r^2 S_{jl}(r) j_l(kr)~, \label{Sjl-til}
\eq
in
analogy to (\ref{Rtil-1}). Since  $ S_{jl}(r)$ also depend on a
parameter called $n_{jl}$,  tabulated in \cite{wave-functions},
the corresponding momentum  RHF functions are:\\
\bqa
 l=0 ~~ &&  \nonumber \\
 n_{j0}=1 & \to &  \wtil S_{j0}(k) = \frac{16 \pi Z_{j0}^{5/2}}{(Z_{j0}^2+k^2)^2}
~~, \nonumber \\
n_{j0}=2 & \to &  \wtil S_{j0}(k) = \frac{16 \pi Z_{j0}^{5/2}
(3 Z_{j0}^2-k^2)}{\sqrt{3}(Z_{j0}^2+k^2)^3} ~~, \nonumber \\
n_{j0}=3 & \to &  \wtil S_{j0}(k) = \frac{64 \sqrt{10} \pi Z_{j0}^{9/2}
( Z_{j0}^2-k^2)}{5 (Z_{j0}^2+k^2)^4} ~~, \nonumber \\
n_{j0}=4 & \to &  \wtil S_{j0}(k) = \frac{64  \pi Z_{j0}^{9/2}
( 5 Z_{j0}^4 - 10 Z_{j0}^2 k^2+k^4)}{\sqrt{35} (Z_{j0}^2+k^2)^5} ~~, \nonumber \\
n_{j0}=5 & \to &  \wtil S_{j0}(k) = \frac{128 \sqrt{14}  \pi Z_{j0}^{13/2}
( 3 Z_{j0}^4 - 10 Z_{j0}^2 k^2+3 k^4)}{21  (Z_{j0}^2+k^2)^6} ~~,
\label{wave-1}  \\[1.cm]
 l=1 ~~ &&  \nonumber \\
 n_{j1}=2 & \to &  \wtil S_{j1}(k) = \frac{64 \pi k Z_{j1}^{7/2}}
 {\sqrt{3} (Z_{j1}^2+k^2)^3}
~~, \nonumber \\
n_{j1}=3 & \to &  \wtil S_{j1}(k) = \frac{64\sqrt{10} \pi k Z_{j1}^{7/2}
(5 Z_{j1}^2-k^2)}{15(Z_{j1}^2+k^2)^4} ~~, \nonumber \\
n_{j1}=4 & \to &  \wtil S_{j1}(k) = \frac{128  \pi k Z_{j1}^{11/2}
( 5 Z_{j1}^2-3 k^2)}{\sqrt{35} (Z_{j1}^2+k^2)^5} ~~, \nonumber \\
n_{j1}=5 & \to &  \wtil S_{j1}(k) = \frac{128 \sqrt{14}  \pi k Z_{j1}^{11/2}
( 35 Z_{j1}^4 - 42 Z_{j1}^2 k^2+3 k^4)}{105  (Z_{j1}^2+k^2)^6} ~~,
\label{wave-2}  \\[1.cm]
 l=2 ~~ &&  \nonumber \\
 n_{j2}=3 & \to &  \wtil S_{j2}(k) = \frac{128 \sqrt{10} \pi k^2 Z_{j2}^{9/2}}
 {5  (Z_{j2}^2+k^2)^4}
~~, \nonumber \\
n_{j2}=4 & \to &  \wtil S_{j2}(k) = \frac{128  \pi k^2 Z_{j2}^{9/2}
(7 Z_{j2}^2-k^2)}{\sqrt{35}(Z_{j2}^2+k^2)^5} ~~. \label{wave-3}
\eqa

The parameters $(C_{jln}, n_{jl}, Z_{jl})$
appearing in (\ref{Rnl-RHF}, \ref{wave-1}-\ref{wave-3}) are  given in
 Tables in \cite{wave-functions}. Parameters $(C_{jln}, ~n_{jl})$ are dimensionless,
while $Z_{jl}$ are expressed   in units of
$m_e \alpha = 3.73~ {\rm keV}$ in \cite{wave-functions}.

We also note that, according to (\ref{wave-1}-\ref{wave-3}, \ref{Rnl-RHF})
and \cite{wave-functions},
the highest  $Z_{j0}$ for each atomic state determines the
relevant range
where the bound electron's momentum
mostly lies\footnote{Compare (\ref{H-function}) and the related variable $Z_H$.}.
In most cases, this  range is found to be
$0\leq  k \lsim 3~ {\rm max}(Z_{j0}) m_e \alpha$.

\clearpage
\newpage

\clearpage
\newpage

\clearpage
\newpage

\begin{figure}[p]
\vspace*{-2cm}
\[
\hspace{-0.5cm}\epsfig{file=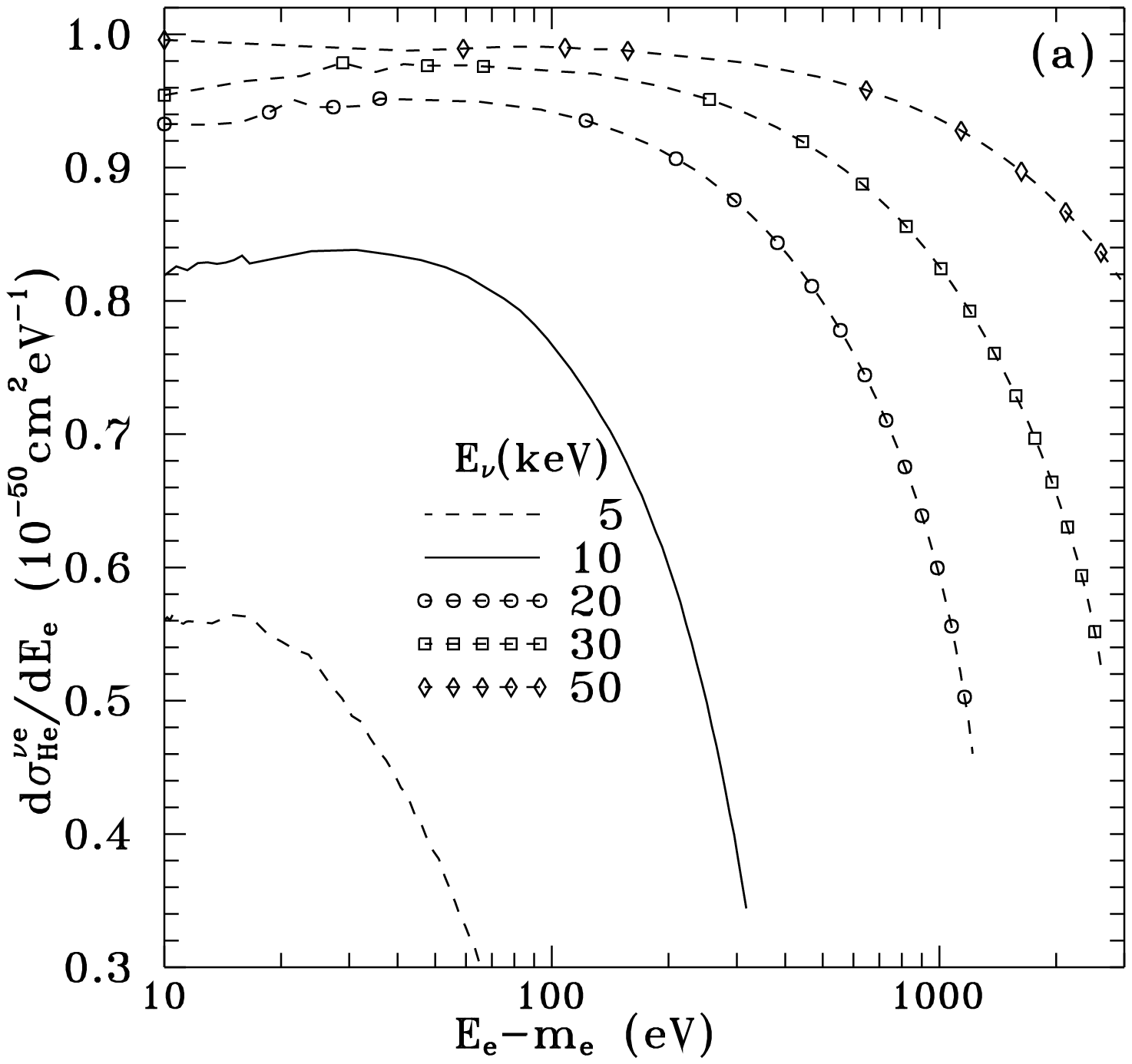,height=7.5cm, width=7.5cm}
\hspace{1.cm}\epsfig{file=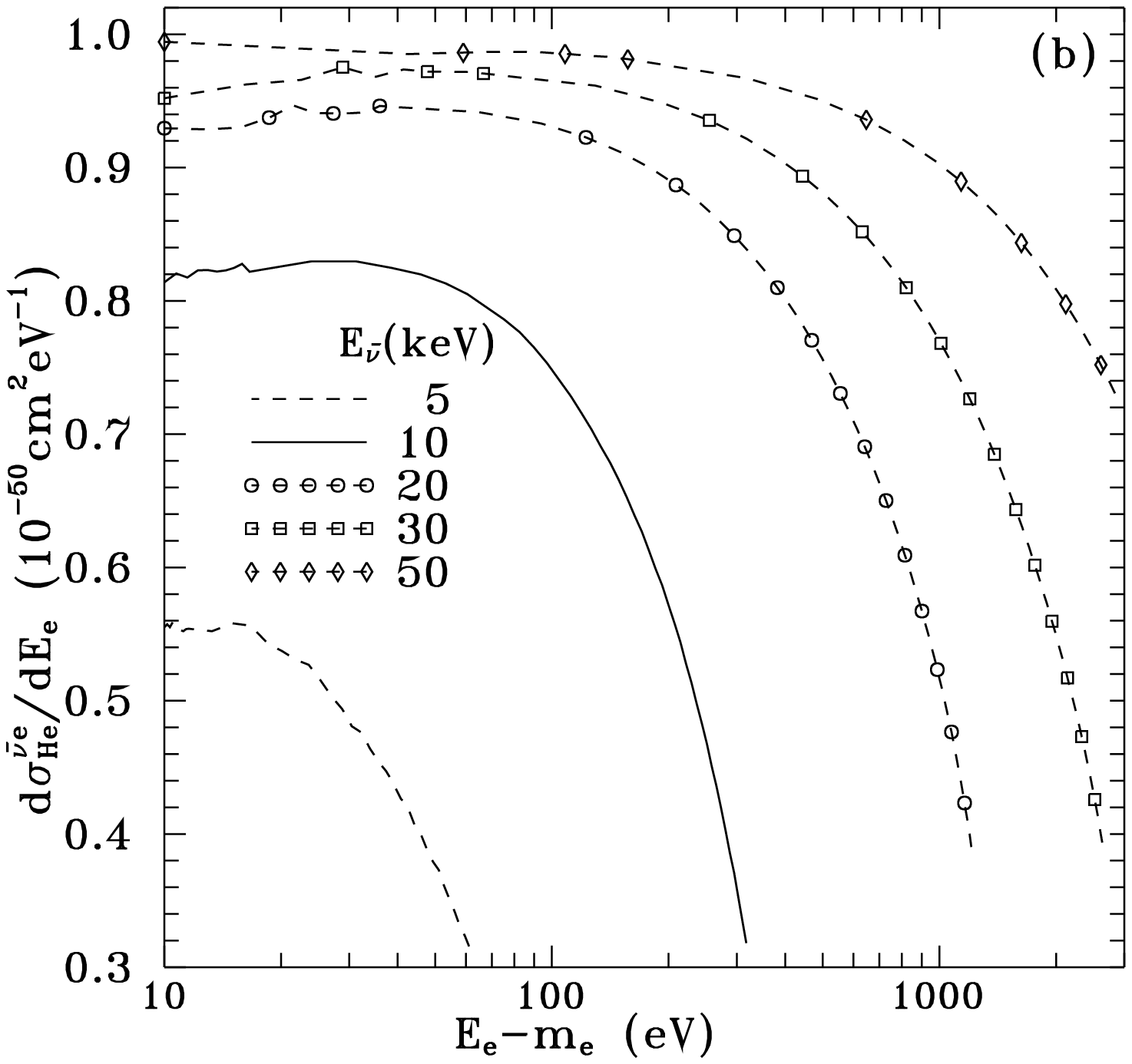,height=7.5cm, width=7.5cm}
\]
\[
\hspace{-0.5cm}\epsfig{file=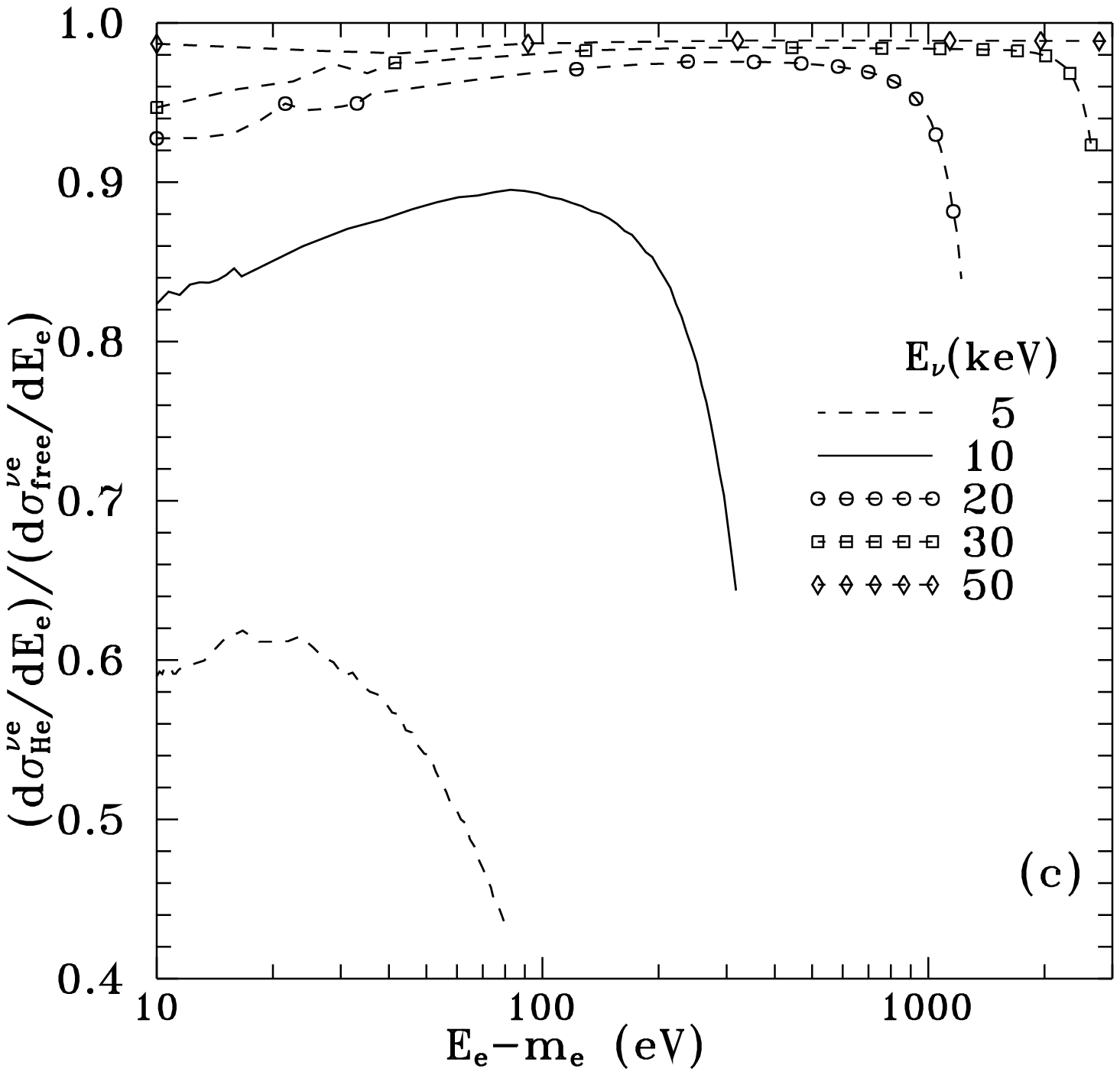,height=7.5cm, width=7.5cm}
\hspace{1.cm}\epsfig{file=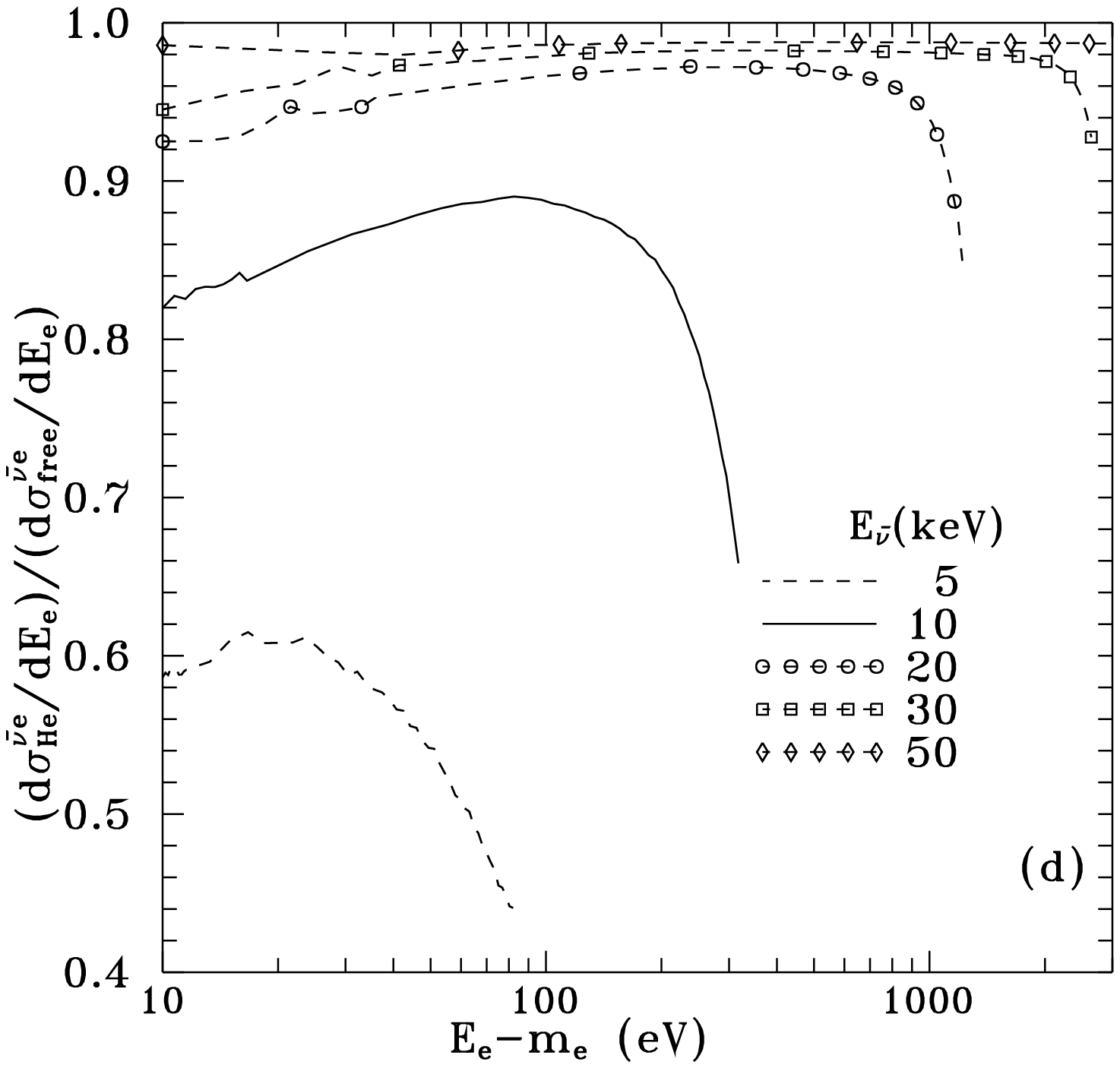,height=7.5cm, width=7.5cm}
\]
\caption[1]{The  electron energy distribution
in   $\nu_e$ (a) or $\bar \nu_e$ (b) ionization of $He$
normalized to one $e^-$ per unit volume, and their
respective ratio (c) and (d)  to
the corresponding  distributions when the initial electron is assumed
as free; see Fig.\ref{Free-energy-fig}.
Variables defined as in caption of Fig.\ref{Free-energy-fig}. }
\label{He-energy-fig}
\end{figure}

\clearpage
\newpage

\begin{figure}[p]
\vspace*{-2cm}
\[
\hspace{-0.5cm}\epsfig{file=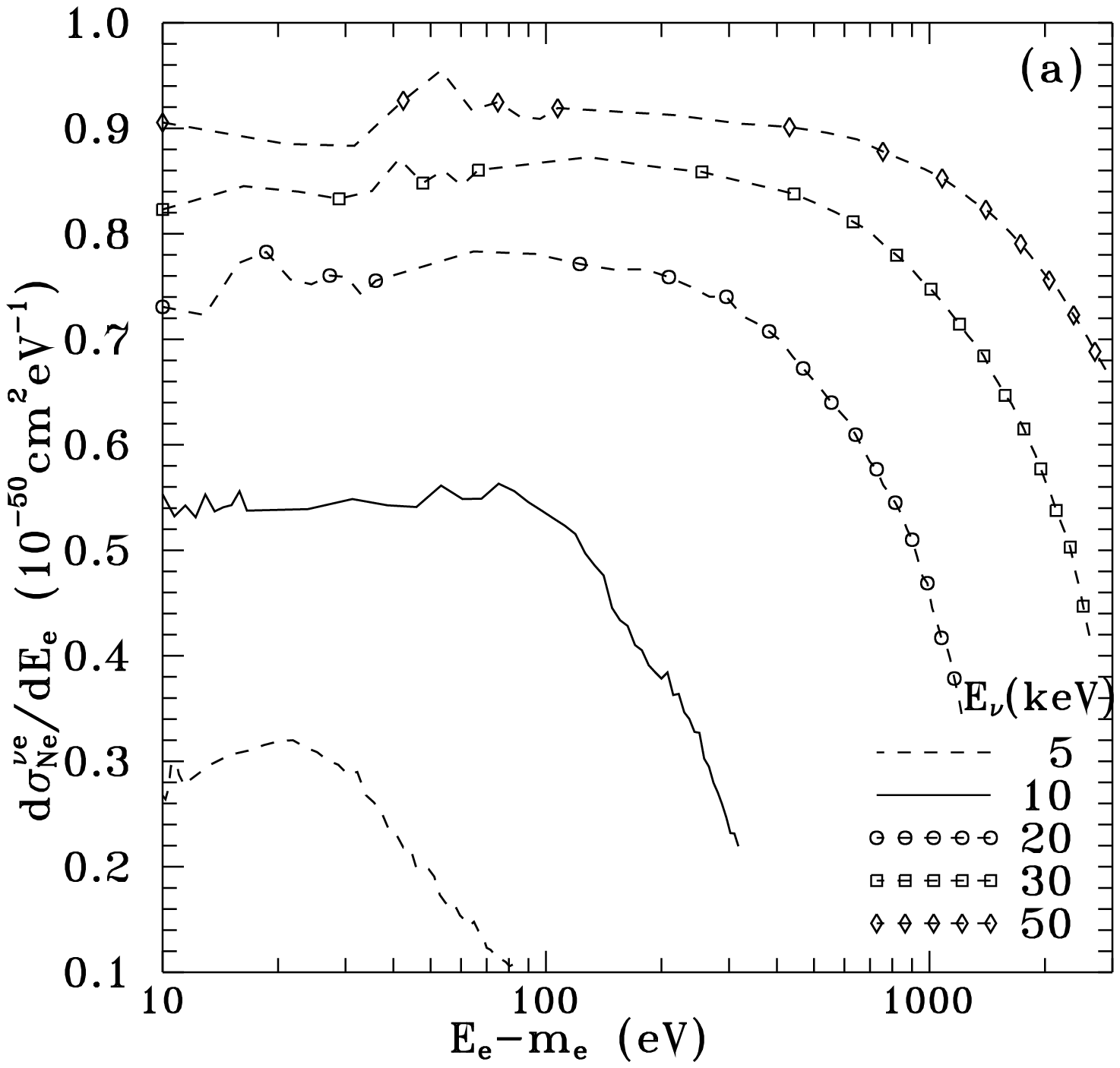,height=7.5cm, width=7.5cm}
\hspace{1.cm}\epsfig{file=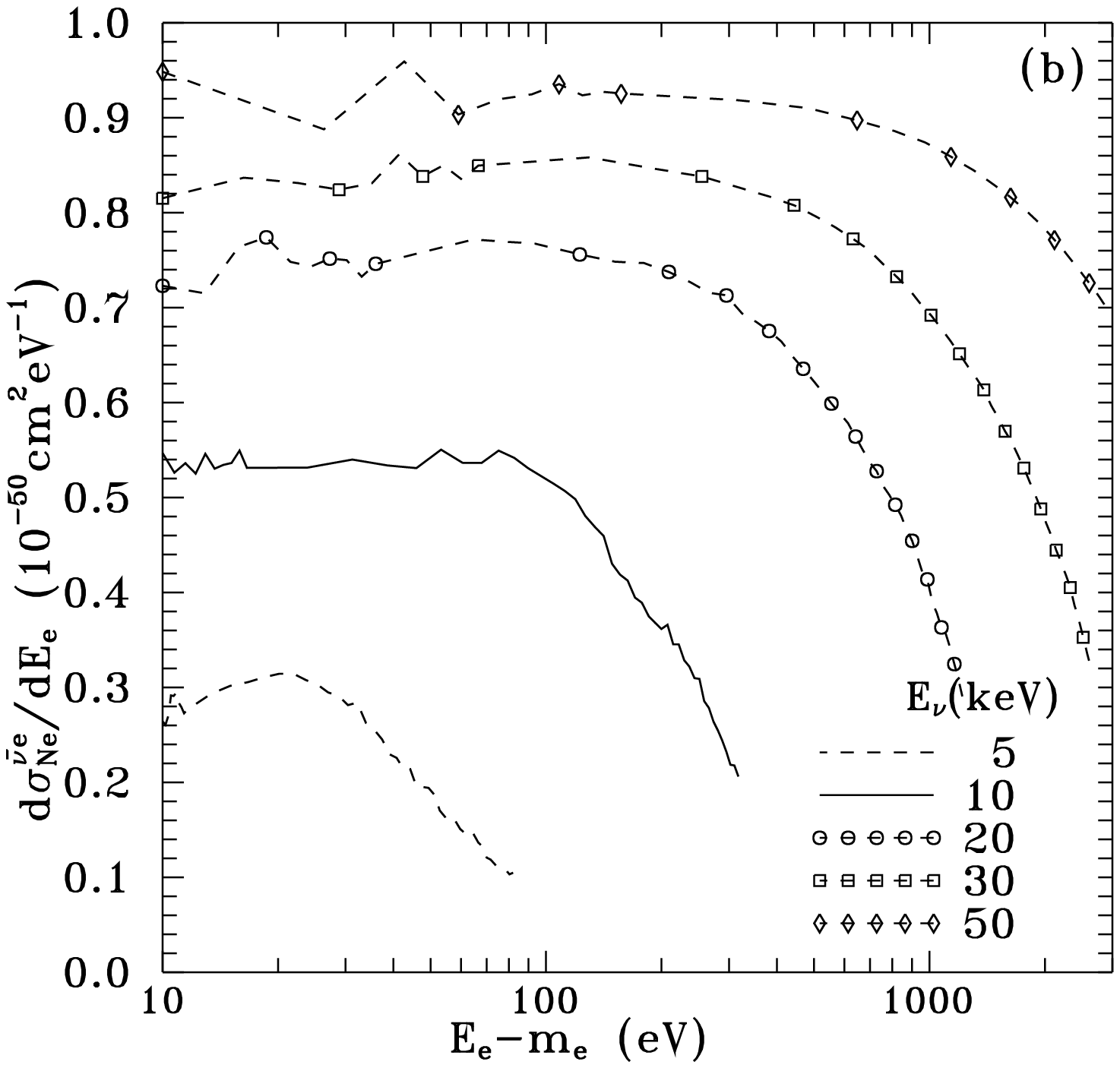,height=7.5cm, width=7.5cm}
\]
\[
\hspace{-0.5cm}\epsfig{file=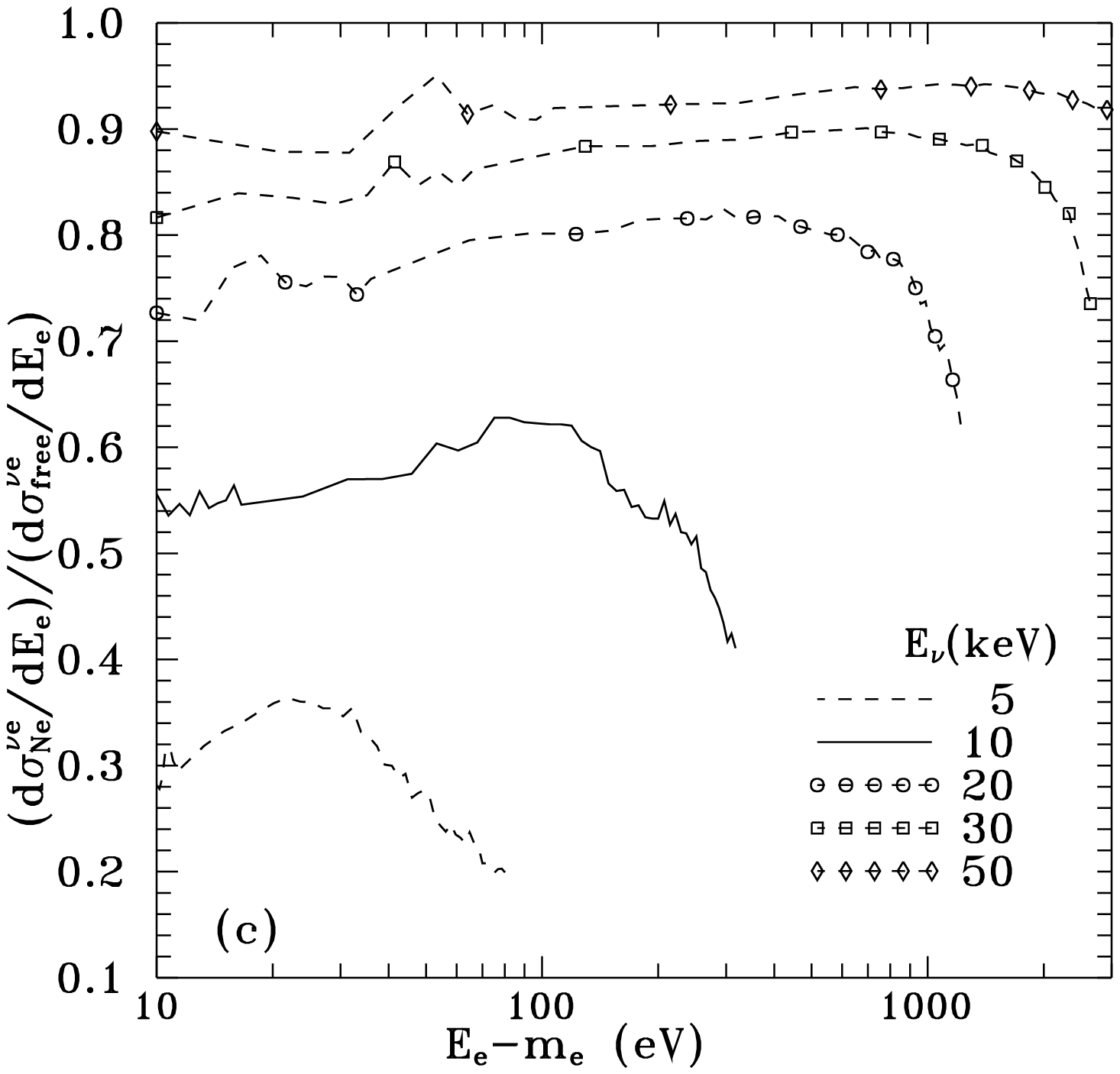,height=7.5cm, width=7.5cm}
\hspace{1.cm}\epsfig{file=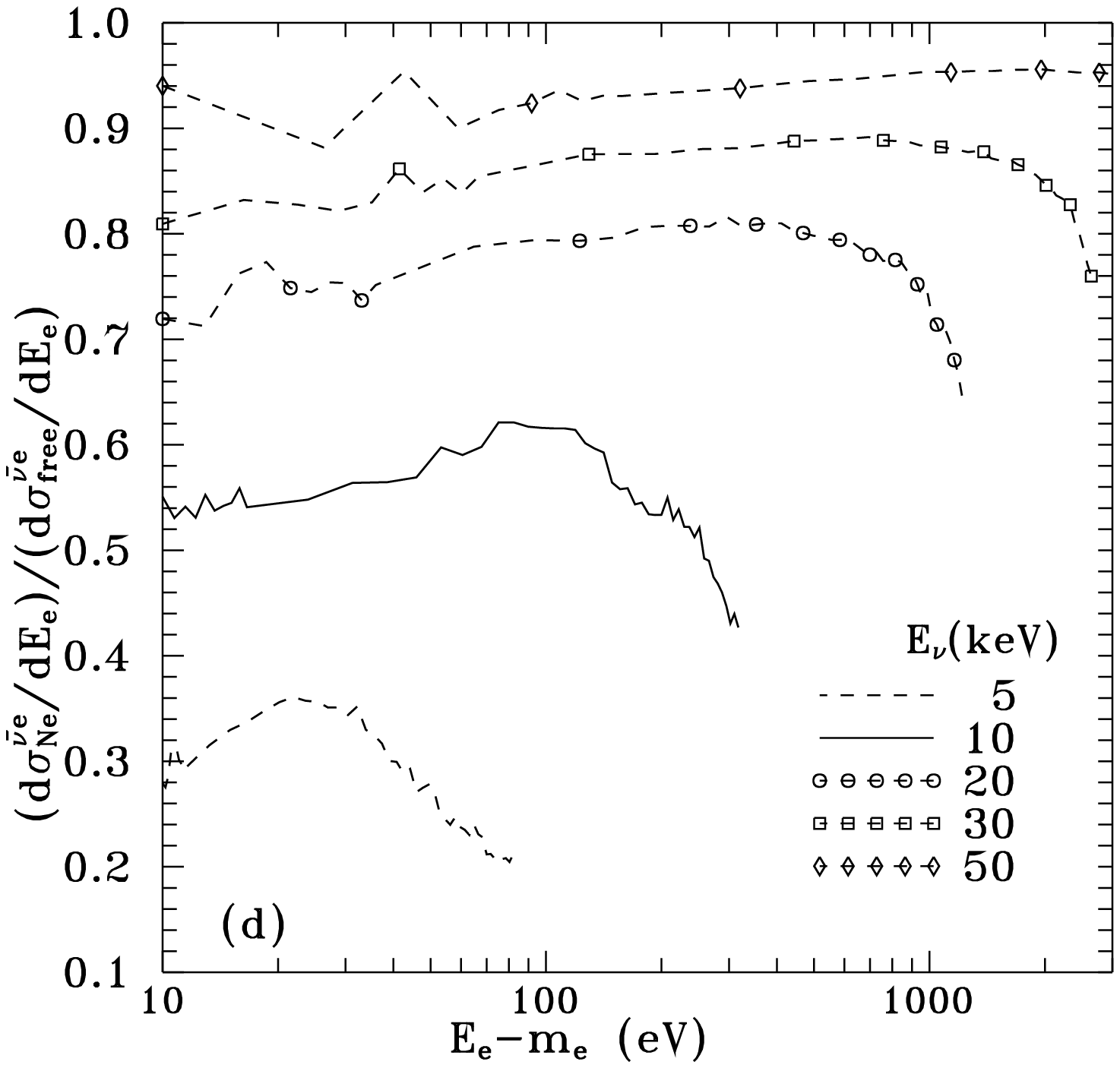,height=7.5cm, width=7.5cm}
\]
\caption[1]{The  electron energy distribution
in   $\nu_e$ (a) or $\bar \nu_e$ (b) ionization of $Ne$
normalized to one $e^-$ per unit volume, and their
respective ratio (c) and (d)  to
the corresponding  distributions when the initial electron is assumed
as free; see Fig.\ref{Free-energy-fig}.
Variables defined as in caption of Fig.\ref{Free-energy-fig}.}
\label{Ne-energy-fig}
\end{figure}

\clearpage
\newpage

\clearpage
\newpage

\clearpage
\newpage

\end{document}